\documentclass[%
 reprint,%
 amsmath,amssymb,
]{revtex4-2}

\usepackage{graphicx}

\usepackage{dcolumn}
\usepackage{bm}
\usepackage{helvet}
\usepackage{setspace}
\makeatletter
\@ifundefined{selectlanguage}{}{%
  \renewcommand\selectlanguage[1]{}%
}
\makeatother

\begin{document}


\title{Granular thermostat implementation within the soft-sphere Discrete Element Method (DEM) framework, considerations and limitations}

\author{Marco Previtali}
 
\affiliation{%
Center for Advanced Engineering Sciences and Technology\\
 Westlake University
}%

\author{Herbert Huppert}
\affiliation{
Faculty of Mathematics\\
 University of Cambridge
}%

\author{Sergio Andres Galindo Torres}%
\email{corresponding author: s.torres@westlake.edu.cn}
\affiliation{%
Center for Advanced Engineering Sciences and Technology\\
Westlake University\\
}%


\date{\today}
\begin{abstract}
Dense granular flows are commonly investigated using the soft-sphere Discrete Element Method (DEM), whereas large-scale applications generally require continuum models with constitutive relations linking stress and deformation to relevant state variables. Granular temperature, defined as the variance of non-affine particle velocity fluctuations, is one such variable and can improve inertial rheological scaling. However, without a constitutive description of its production, transport and dissipation, temperature-based relationships remain of limited predictive value. Moreover, experimental forcing such as vibration changes temperature together with velocity statistics, correlations and other aspects of the granular state, making their individual effects difficult to isolate.

An alternative is to use a thermostat algorithm to control the fluctuation energy directly. Although thermostat algorithms are well established in Molecular Dynamics (MD), their behaviour in dissipative DEM systems has received comparatively little attention. We show that conventional Langevin and Nos\'e-Hoover thermostats exhibit complementary limitations: Langevin control requires strong coupling to offset collisional dissipation, thereby damping particle dynamics, whereas Nos\'e-Hoover does not independently disrupt the correlations and segregation generated by repeated inelastic collisions, leading to non-ergodic and numerically unstable states.

To address these limitations, we introduce two pairwise hybrid formulations combining deterministic temperature regulation with stochastic decorrelation. Both enforce the prescribed temperature, while variation of the stochastic decorrelation timescale modifies velocity statistics and associated mesoscopic properties such as diffusivity. The two formulations nevertheless retain distinct statistical signatures, demonstrating that fluctuation magnitude, decorrelation and forcing statistics represent separate aspects of the granular state. Finally, application to pressure-controlled simple shear shows that increasing granular temperature at fixed inertial number reduces the apparent friction, consistent with previously reported trends. The framework therefore provides controlled reference states for comparing differently forced granular systems and identifying the variables required for temperature-dependent constitutive models.
\end{abstract}
\maketitle
\section{Introduction}
Granular materials occur in a wide range of natural and industrial environments. Under quasi-static conditions, their mesoscopic response is governed primarily by frictional contact networks, whereas increasing inertial effects progressively disrupt these force chains and induce a fluid-like state \citep{vescovi2016merging,mandal2021rheology}. Soft-sphere Discrete Element Method (DEM) simulations are widely used to investigate this range of behaviour because they explicitly resolve finite-duration frictional contacts. This distinguishes them from hard-sphere or event-driven collision models, in which interactions are treated as instantaneous binary collisions and are therefore better suited to describe dilute regimes \citep{cundall1979discrete}.\\
\indent Despite their ability to resolve particle-scale mechanisms, DEM simulations remain computationally expensive, limiting their direct application to many engineering and geophysical-scale problems. This motivates the development of continuum descriptions in which constitutive relations are used to connect macroscopic fields such as stress, density and strain rate. For dense granular flows, one of the most widely used such descriptions is the $\mu(I)$ rheology introduced by \cite{jop2006constitutive}, which relates the apparent friction $\mu=\tau/p$ to the inertial number $I=\dot{\gamma}d\sqrt{\rho_p/p}$, where $\tau$ is the shear stress, $p$ the confining stress, $\dot{\gamma}$ the strain rate, $d$ the particle diameter and $\rho_p$ the particle density.\\
\indent A description based on $I$ alone is nevertheless incomplete, as different boundary-value problems can exhibit different rheological responses at comparable inertial conditions \citep{gdr_midi_dense_2004,kamrin_nonlocal_2012,dumont2023microscopic}. Granular temperature, defined here from the variance of the non-affine particle velocity fluctuations, provides an additional measure of the kinetic state \citep{alessio2026dense}. Incorporating temperature can improve rheological collapse: \cite{kim_power-law_2020}, for example, proposed $\mu\Theta^{1/6}\propto I$, where $\Theta=\rho_pT/p$, which has subsequently been applied to vibrated shear \citep{irmer_granular_2024}. The scaling is not universal, however, and its dependence on properties such as inter-particle friction has been demonstrated \citep{man_friction-dependent_2022}. More fundamentally, a relation of the form $\mu(I,\Theta)$ is not predictive by itself because it does not describe the production, transport and dissipation of $T$ \citep{pouliquen2026non,alessio2026dense}.\\
\indent Extended kinetic theory provides an energetic framework for describing these processes by relating mass, momentum and fluctuation-energy transport to particle-scale collision statistics \citep{berzi_granular_2024}. Its application to dense granular flows nevertheless requires closures for quantities such as the radial distribution function at contact and pre-collisional velocity correlations. The latter are particularly important because classical kinetic theory assumes molecular chaos, whereby particles approaching a collision have statistically uncorrelated velocities. Dense and inelastic granular systems can instead develop persistent velocity correlations and non-Gaussian velocity distributions, modifying collisional dissipation and transport \citep{goldhirsch2008introduction,garzo1999dense,garzo2004diffusion}. Closure relations calibrated under specific conditions, such as homogeneous simple shear, therefore incorporate the relationships between temperature, density, shear and velocity correlations that emerge under that forcing condition \citep{berzi2014extended}; their transferability to different mechanisms of energy injection cannot be assumed a priori.\\
\indent This presents a significant problem for experiments and simulations used to inform granular rheology. Conventional methods of changing the kinetic state do not vary temperature alone: rough or vibrating boundaries simultaneously modify slip, shear localisation, dilation, anisotropy, velocity distributions and correlations \citep{wang2025basal,prevost2002forcing,fei2026scaling}. Consequently, changes in rheology or transport cannot generally be attributed uniquely to scalar granular temperature, and systems with the same $T$ may retain different kinetic and structural states.\\
\indent A different approach, widely used in Molecular Dynamics (MD), is to employ a thermostat algorithm that adds or removes energy at the particle level to maintain a prescribed kinetic temperature. Although such algorithms can in principle be applied to soft-sphere DEM \citep{cundall1979discrete}, their behaviour in dissipative soft-sphere granular systems has received comparatively little attention. Most conventional thermostat formulations were developed for molecular systems in which persistent collisional dissipation, friction, clustering and velocity correlations do not play the same role as in granular DEM, and these processes can substantially modify their response.\\
\indent This paper addresses this gap by investigating conventional thermostat algorithms in soft-sphere DEM and developing hybrid formulations suited to dissipative granular systems. The objective is not to reproduce a particular vibrating or rough boundary, but to construct controlled reference states in which the magnitude of velocity fluctuations can be prescribed separately, to some extent, from the statistical character of the forcing that generates them. This provides a means of determining which transport and rheological properties are governed primarily by scalar granular temperature and which additionally depend on velocity statistics, correlations and particle structure. The comparison with experimentally or naturally forced systems is therefore intended to be state-based rather than based on a one-to-one mapping between thermostat and apparatus parameters: systems may first be matched in $I$ and $T$, and then compared through additional measurable descriptors such as velocity statistics, correlation lengths and near-contact structure.\\
\indent The paper is structured as follows. Existing deterministic and stochastic thermostat algorithms are first reviewed and their limitations in inelastic granular systems are examined. Two hybrid formulations are then introduced, followed by parametric studies of thermostat formulation, stochastic forcing, interaction parameters and collisional inelasticity. Finally, the thermostat is applied to pressure-controlled simple shear to demonstrate controlled exploration of the $(I,T)$ state space.
\section{Thermostat algorithms}
\subsection{State of the art}\label{sc_state_art}

A non-exhaustive overview of existing thermostat implementations is given in the following, broadly subdivided into deterministic and stochastic approaches.\\ 
\indent Deterministic thermostats modify the kinetic state of the system through feedback from the instantaneous system temperature. An example of this is the Berendsen thermostat, which periodically (e.g. every 100-1000 steps) rescales particle velocities to match the desired instantaneous temperature \citep{ruiz2018effect}. However, as discussed in \cite{braun2018anomalous}, simple velocity rescaling violates the equipartition theorem and systematically redistributes energy from high (vibrational) to low frequency (translational) modes, leading to the so-called \textit{flying ice cube} effect, i.e. particle clusters moving coherently without internal interactions \citep{basconi_effects_2013}. \cite{ruiz2018effect} suggests mitigating this effect by employing thermostats that respect the canonical distribution of kinetic energy, such as \cite{bussi_canonical_2007}. The Nosé-Hoover thermostat \citep{nose1984unified} is perhaps the most widely adopted deterministic method: it introduces an additional degree of freedom into the Hamiltonian, which acts as a dynamic friction coefficient, providing smooth and continuous temperature control. However, this formulation is also known to divert the kinetic energy of the system toward specific modes, producing non-ergodic behaviour. To mitigate this issue, \cite{martyna_nosehoover_1992} substitutes the single degree of freedom of the original Nos\'e-Hoover thermostat with a series of coupled reservoirs. This chain of thermostats acts as a buffer between the measured temperature difference and the global friction parameter, promoting a more uniform redistribution of energy across all modes. Despite this improvement, Nos\'e-Hoover remains a fully deterministic approach, and therefore always exhibits some degree of non-ergodicity \citep{patra_nonergodicity_2014}.\\
\indent Stochastic thermostats update the kinetic state of each particle by sampling a prescribed statistical distribution. Because the corrections are independent of each other and from the collective state of the system, these algorithms are also called local. A simple implementation of this concept is represented by the Andersen thermostat, which simulates random collisions with a fictitious heat bath by intermittently assigning new momenta sampled from the Maxwell-Boltzmann distribution. However, the collision frequency is a non-physical parameter which strongly affects the behaviour of the system: if it is too high, it restricts diffusion and natural dynamics; if it is too low, it can lead to velocity segregation and loss of ergodicity \citep{verbeek_advantages_2022}. The Langevin thermostat mitigates these effects by applying a continuous correction and adding a dissipative term alongside the stochastic noise. This additional term restores energy balance at equilibrium, ensuring that the Langevin equation obeys the fluctuation-dissipation theorem \citep{langevin1908theory}. Doing so, it formally connects the stochastic behaviour of an individual particle (Langevin dynamics) to the ensemble behaviour described by the Fokker-Planck equation \citep{phan2013understanding}. However, the combination of uncorrelated stochastic forces and velocity damping has a significant dissipative effect on the underlying dynamics of the system. When a large coupling value is employed, the thermostat suppresses random-walk behaviour: the particles vibrate at the target temperature, but remain stationary without diffusing according to Brownian motion \citep{basconi_effects_2013}. On the other hand, deterministic thermostats are generally considered to provide a more realistic representation of the underlying dynamics of the system, but they can still affect the emerging properties at the meso-scale \citep{li2019influence,halonen_further_2023}.\\
\indent The trade-off between thermostats that preserve the underlying dynamics and those that respect the canonical distribution led to the development of hybrid approaches. Observing that deterministic and stochastic formulations have opposite effects on the rheological properties of molecular fluids, \cite{stoyanov_molecular_2005} proposed a probabilistic combination in which interactions are treated either deterministically or stochastically. \cite{bussi_canonical_2007} combined deterministic temperature control with stochastic rescaling to recover the canonical distribution, while \cite{leimkuhler2009gentle} introduced stochasticity into a Hoover-style feedback variable, improving ergodicity and numerical stability \citep{leimkuhler2015numerical,leimkuhler2016pairwise}. These approaches motivate the combination of deterministic temperature regulation and stochastic decorrelation considered below.

The exact particle-level forcing operators used by numerical thermostats do not have direct experimental counterparts, although several experiments approach similar classes of energy-transfer statistics. Vibrated granular systems provide the most direct examples. Changing a substrate from flat to rough modifies both velocity correlations and velocity distributions \citep{prevost2002forcing}, while sidewall dissipation can modify the spatial temperature field, velocity statistics and convection throughout the bulk \citep{windows2013thermal}. Randomised forcing can move the system closer to stochastic-heating assumptions: \cite{reis2007forcing} investigated a granular fluid in relation to stochastic thermostatting, while \cite{windows2013boltzmann} showed that randomising energy transfer through a mobile-particle base produces more Gaussian velocity statistics, greater temperature isotropy and a closer approximation to molecular chaos. Spatially distributed forcing has also been realised through the magnetic \textit{bulk thermostat} of \cite{adachi2019magnetic}, which injects fluctuation energy throughout the particle ensemble without relying on a mechanically vibrating boundary. These systems provide experimental counterparts to different classes of energy injection, although none maps uniquely onto a numerical thermostat because physical forcing simultaneously introduces additional effects on the resulting granular state.

\subsection{Application to granular media}\label{sc_granular_requirements}

Applying a thermostat to a driven granular system introduces two requirements that are less restrictive in conventional MD. First, the thermostat must distinguish random velocity fluctuations from the mean flow. Second, it must maintain a steady state despite the continuous loss of fluctuation energy through inelastic and frictional interactions.

\subsubsection{Galilean Invariance}
Any thermostat used in hydrodynamic simulations requires Galilean invariance: velocity fluctuations only exist relative to an inertial frame of reference, which is generally assumed static for most MD simulations. For driven systems, a suitable frame of reference can be obtained for each particle by sampling the local group velocity, e.g. through spatial kernel functions. This represents a straightforward solution, but it scales poorly and becomes computationally prohibitive for large simulations. A more efficient approach involves the definition of a pairwise frame of reference, as done through the Dissipative Particle Dynamics (DPD) framework by \cite{hoogerbrugge1992simulating}. While the method detailed in the original paper also possesses other characteristics (e.g. coarse-grained particle representation), the main intuition behind DPD is that the relative velocity of two particles at rest within the same inertial frame of reference is zero, and any deviation can be quantified as a thermal fluctuation. By the same logic, if a thermostat modifies the relative velocity of a particle pair, it does not affect their average (group) velocity, retrieving Galilean invariance. Still, as with any non-local approach, DPD requires the definition of a characteristic length, the interaction range ($r_t$), which can introduce various non-physical artifacts: a large $r_t$ value smooths the velocity field, while a small one prevents the homogeneous application of the thermostat, with the result that a large portion of the particles is ignored and the target temperature is not correctly enforced \citep{phan2013understanding}. In the Lowe-Andersen thermostat, a momentum conserving and Galilean invariant analogue of the Andersen approach, the $r_t$ controls the fluid viscosity \citep{koopman2006advantages}. \cite{verbeek_advantages_2022} notes that $r_t$ can act as a proxy for the thermostat application frequency: if the value is too large, any given particle participates to many thermostat pairs, causing the algorithm to be applied to it multiple times within a short time frame. When combined with stochastic temperature control, this reproduces the same over-dampened dynamics obtained for excessive thermostat coupling strength. 
The appropriate thermostat interaction range is investigated explicitly in section \ref{sc_results}.

\subsubsection{Non-equilibrium systems}

Granular systems are inherently dissipative and therefore require continuous energy input to maintain a non-zero stationary temperature. As a first approximation, consider a homogeneous system of monodisperse smooth particles in which collisional cooling can be represented as an effective drift in velocity space, $\langle v\rangle\rightarrow0$. Inelastic collisions continuously remove kinetic energy and violate detailed balance, so the system does not possess the equilibrium energy balance assumed in conventional molecular thermostatting. Thermostats based on intermittent state resets, such as simple velocity rescaling, are therefore poorly suited to the continuously driven non-equilibrium steady states considered here. We instead focus on force-based Langevin (LA) and Nos\'e-Hoover (NH) formulations, which act continuously. Consider first a homogeneous cooling system coupled to a Langevin thermostat,
\begin{equation}\label{eq_langevin_maintext}
dv=-\left(\Gamma+\xi_\text{La}\right)v\,dt+\sqrt{2\xi_\text{La}\hat{T}}\,dW,
\end{equation}
where $\xi_\text{La}$ is the thermostat coupling strength, $\hat{T}$ is the prescribed temperature and $W(t)$ is a standard Wiener process, such that $\langle dW\rangle=0$ and $\langle dW^2\rangle=dt$. The parameter $\Gamma$ represents homogeneous collisional cooling and is assumed constant for $T\simeq\hat{T}$. The corresponding Fokker-Planck equation is
\begin{equation}\label{eq_fokker_planck_langevin}
\frac{\partial P}{\partial t}=\frac{\partial}{\partial v}\left[(\Gamma+\xi_\text{La})vP\right]+\xi_\text{La}\hat{T}\frac{\partial^2P}{\partial v^2}.
\end{equation}
Setting $\partial P/\partial t=0$ and imposing $P\rightarrow0$ as $|v|\rightarrow\infty$ gives the stationary Gaussian distribution
\begin{equation}
P(v)=\sqrt{\frac{\Gamma+\xi_\text{La}}{2\pi\xi_\text{La}\hat{T}}}\exp\left[-\frac{\Gamma+\xi_\text{La}}{2\xi_\text{La}\hat{T}}v^2\right],
\end{equation}
with variance
\begin{equation}\label{eq_temperature_scaling_langevin}
T=\langle v^2\rangle={\xi_\text{La}\hat{T}}/({\Gamma+\xi_\text{La}}).
\end{equation}
Therefore, although the stationary distribution remains Gaussian under this approximation, the prescribed temperature is recovered only in the strong-coupling limit $\xi_\text{La}\gg\Gamma$. Langevin temperature control therefore requires increasingly strong damping as collisional dissipation increases, which can substantially alter the underlying particle dynamics \citep{halonen_further_2023}.\\
\indent Deterministic thermostats avoid this explicit stochastic damping and can preserve molecular-fluid dynamics \citep{basconi_effects_2013}. Consider a dissipative system coupled to a Nos\'e-Hoover thermostat,
\begin{equation}\label{eq_nose_hoover_partial_form}
\begin{aligned}
dv&=-(\Gamma+\zeta)v\,dt,\\
d\zeta&=(v^2-\hat{T})\,dt/(Q\hat{T}),
\end{aligned}
\end{equation}
where $\zeta$ is the additional thermostat variable and $Q=\xi_\text{NH}^2$ is its thermal inertia. Unlike Langevin dynamics, eq. (\ref{eq_nose_hoover_partial_form}) contains no explicit diffusive term; instead, $\zeta$ evolves in response to the instantaneous kinetic energy. Under the homogeneous constant-cooling approximation, the stationary Liouville equation derived in section \ref{sc_nosehoover_Pv} gives
\begin{equation}
P(v)=\frac{1}{\sqrt{2\pi\hat{T}}}\exp\left[-\frac{v^2}{2\hat{T}}\right],
\end{equation}
and therefore $\langle v^2\rangle=\hat{T}$. Within this approximation, Nos\'e-Hoover can thus maintain the prescribed variance without the strong-coupling requirement of the Langevin thermostat. \\
\indent The homogeneous approximation does not, however, account for the instability of freely cooling inelastic granular systems. A local increase in particle density produces an increment in the local collision rate and therefore the local dissipation rate. This reduction in relative velocities makes particles more likely to remain in close proximity, further increasing the local density. This clustering instability, typically referred to as collisional cooling, reverses the homogenising effect of diffusion typical of Brownian motion and drives the transition from the homogeneous cooling state (HCS) toward an inhomogeneous cooling state (ICS) \citep{goldhirsch2008introduction}. The accompanying spatial and velocity correlations lead to departures from the Maxwellian velocity statistics assumed above and, at sufficiently strong clustering, can produce locally dense or solid-like regions \citep{goldhirsch2008introduction,komatsu2015roles}.\\
\indent This transition has been investigated analytically through Sonine expansions of the Enskog-Boltzmann equation \citep{garzo1999dense} and numerically through Direct Simulation Monte Carlo and molecular dynamics \citep{montanero2000computer,garzo2004diffusion}. Increasing inelasticity promotes spatial and pre-collisional velocity correlations and progressively shifts the velocity distribution away from its Gaussian limit \citep{brilliantov2001granular,brilliantov2001granularb,garzo2004diffusion}. A global Nos\'e-Hoover thermostat controls the average temperature but provides no independent mechanism for disrupting these correlations. As demonstrated in section \ref{sc_results}, the resulting clustered and velocity-segregated states can also become numerically problematic in soft-sphere DEM, where large velocity fluctuations interact with stiff contact laws and explicit time integration \citep{fullmer2022divergence}.

\subsubsection{Toward a hybrid thermostat}
The addition of a stochastic term provides a direct means of disrupting the correlations generated by repeated inelastic collisions. The strong-coupling requirement of pure Langevin temperature control, $\xi_\text{La}\gg\Gamma$, can be avoided by retaining Nos\'e-Hoover as the primary temperature-control mechanism. Such stochastic forcing also has a useful physical interpretation. In kinetic descriptions of heated granular gases, random forcing has long been used as an idealised homogeneous energy source \citep{garzo2002transport}. Granular suspensions have similarly been modelled by representing interactions with an interstitial gas through viscous drag and a stochastic Langevin-like force \citep{garzo2007enskog,gomez2023diffusion}. \cite{gonzalez2022kinetic} instead considered explicit elastic collisions between granular particles and a surrounding molecular gas maintained at fixed temperature, showing that the corresponding collision operator reduces to a Fokker-Planck, Langevin-like description in the Brownian limit. The conventional Langevin thermostat can therefore be interpreted as an idealised homogeneous heat bath, in which damping and stochastic forcing represent the cumulative effect of interactions with an external reservoir.\\
\indent Hybrid deterministic-stochastic thermostats have previously been introduced to improve ergodicity in molecular systems \citep{stoyanov_molecular_2005,leimkuhler2015numerical}. Here, the stochastic contribution is instead applied directly to particle motion to suppress collisional correlations, while Nos\'e-Hoover regulates the prescribed temperature. The most straightforward implementation is to add a conventional Langevin term to a Nos\'e-Hoover thermostat. We refer to this formulation as \textit{Naive Addition} (NA), since the two existing formulations are combined without modification. Assuming an approximately homogeneous stationary state, the average collisional cooling rate can be represented by $\Gamma$ \citep{garzo2002transport}, giving
\begin{eqnarray}\label{eq_combined_maintext}
dv
&=&-(\Gamma+\xi_\text{La}+\zeta)v\,dt
+\sqrt{2\xi_\text{La}\hat{T}}\,dW,
\nonumber\\
d\zeta
&=&(v^2-\hat{T})\,dt/Q.
\end{eqnarray}
As shown in section \ref{sc_naive_addition_Pv}, under the homogeneous stationary approximation the Langevin damping and fluctuation terms cancel out, so that the velocity distribution retains the target variance imposed by Nos\'e-Hoover. The two components therefore have approximately distinct roles: Nos\'e-Hoover controls the mean fluctuation energy, whereas $\xi_\text{La}$ controls stochastic decorrelation. This separation is not exact in an inelastic granular system because stochastic forcing also modifies collision statistics, dissipation and particle diffusion \citep{garzo2002transport,visco2007power}. The full Langevin term nevertheless retains the explicit damping contribution $-\xi_\text{La}v$. This term is required by the fluctuation-dissipation relation when Langevin dynamics represent an equilibrium heat bath, but is unnecessary for global temperature regulation here because that function is already performed by Nos\'e-Hoover. Retaining it therefore introduces an additional suppression of the underlying particle dynamics.\\
\indent This motivates a second formulation, termed \textit{Scaled Langevin} (SL), in which the Langevin damping is removed and the stochastic amplitude is scaled according to the instantaneous kinetic state of the interacting pair. To quantify this state, we use the normal relative velocity $v_n$ introduced by the DPD-based pairwise implementation described in section \ref{sc_implementation}, and define the corresponding pair temperature as $T_p=v_n^2/2$. The one-dimensional dynamics can then be written as
\begin{equation}\label{eq_SL_1D}
dv_n=-(\Gamma+\zeta)v_n\,dt+\sqrt{2\xi_\text{La}(\hat{T}-T_p)_+}\,dW,
\end{equation}
where $(x)_+=\max(x,0)$. The stochastic forcing therefore decreases as $T_p$ approaches the target and vanishes for $T_p\geq\hat{T}$. Unlike the additive noise of the standard Langevin and NA formulations, SL introduces multiplicative, piecewise stochastic forcing whose diffusion coefficient depends on $v_n$. The Gaussian stationary treatment used for the preceding formulations therefore does not directly carry over to SL, and its stationary velocity distribution is only characterised numerically in section \ref{sc_results}.\\
\indent However, this state dependence provides a qualitative analogy between SL and the mechanically vibrating boundaries commonly used to heat granular systems \citep{plati2021getting,irmer_granular_2024}, which differentiate it from standard white-noise forcing. A vibrating wall possesses a finite characteristic velocity $V_0=A\omega$, where $A$ is the vibration amplitude and $\omega$ its angular frequency. The mean energy transferred during particle-wall collisions decreases as the incoming particle velocity increases relative to this scale and can become negative at sufficiently large velocities \citep{mcnamara1997energy}. As derived in section \ref{sc_SL_vibrating_wall}, SL represents an idealised version of this behaviour by progressively reducing the stochastic energy input and suppressing it above a finite kinetic scale.
\section{Methods}
\subsection{Model setup}
The different thermostat formulations are tested in a granular system at rest ($\boldsymbol{v}_g=0$) with a baseline temperature of zero. To prevent crystallisation, a quasi-monodisperse particle size distribution is used. The diameter $d$ of a given particle is calculated as $d = d_{min}d_{max}/[d_{max}-R(d_{max}-d_{min})]$, where $R$ is a random number sampled from a uniform distribution between 0 and 1 and $d_{min}$ and $d_{max}$ are the minimum and maximum particle diameters, respectively. This formulation is chosen so that all particle masses are equally represented \citep{poschel2005computational}. The particles are placed on a regular grid (spacing = $d_{max}$) within a periodic cubic domain with a side size $L= 20 d_{max}$. After generation, 40\% of the particles in the domain are randomly removed, resulting in a solid fraction $\phi\approx0.27$. A random initial velocity (from 0 to 0.01) is assigned to the particles to enable Nos\'e-Hoover temperature control. Subsequently, the thermostat is activated and the simulation runs an arbitrary large period of time $t_f$ (= 50). Particle interactions follow a standard Soft-Sphere DEM approach: the model allows for small particle overlaps, which are then used to calculate the repulsive force according to the prescribed spring model. For simplicity, we adopt a linear elastic (Hookean) contact law, with a viscous dashpot in the normal direction and Coulomb frictional slider in the tangential direction. The force increment for a contact between two particles with subscripts $i,j$ is $\boldsymbol{F}_{ij}=K_n\delta\boldsymbol{n}_{ij}+K_t\delta \boldsymbol{s}_{ij}-\boldsymbol{n}\gamma_n v_n$, where $K_{n,t}$ are the normal and tangential stiffness coefficients, $\delta\boldsymbol{n}_{ij}$ and $\delta\boldsymbol{s}_{ij}$ are the overlap increments in the normal and tangential direction (the latter is capped using the Coulomb model), and the last term introduces energy loss in the normal direction due to inelasticity ($\boldsymbol{n}$ is the normal vector of the contact). The damping coefficient $\gamma_n$ is obtained from the coefficient of restitution ($e$) according to  \cite{brilliantov1996model}:
$\gamma_n=\sqrt{{[4m_rK_n(\log(e))^2]/[\pi^2+(\log(e))^2]}}$. All relevant parameters are listed in Table \ref{tb_dem_parameters}. No specific system of units is adopted and parameters are expressed in terms of their fundamental dimensions: length $L$, mass $M$ and time $T$. From there, force is expressed as $MLT^{-2}$ and temperature as $L^2T^{-2}$.

\begin{table}[h]
\caption{\label{tb_dem_parameters}List of relevant DEM parameters. Dimensions: length $L$, mass $M$, time $T$. Force $F$ is expressed as MLT$^{-2}$.}
\begin{ruledtabular}
\begin{tabular}{lccc}
\textrm{Property}&
\textrm{Symbol}&
\textrm{Value}&
\textrm{Dimensions}\\
\colrule
Particle diameter (maximum) & $d_{max}$ & 0.2 & L \\
Particle diameter ratio & $d_{min}/d_{max}$ & 0.9 & $-$\\
Particle density & $\rho_p$ & 3 & ML$^{-3}$\\
Solid fraction & $\phi$ & 0.27 & $-$\\

Normal stiffness & $K_n$ & 2E+5 & MT$^{-2}$ \\
Tangential stiffness & $K_t$ & 2E+5 & MT$^{-2}$\\
Restitution coefficient & $e$ & 0.4 & $-$ \\
Tangential friction coefficient & $\mu_p$ & 0.3 & $-$\\
\end{tabular}
\end{ruledtabular}
\end{table}

\subsection{Implementation details}\label{sc_implementation}
\begin{figure*}[!htb]
    \centering
    \includegraphics[width=\linewidth]{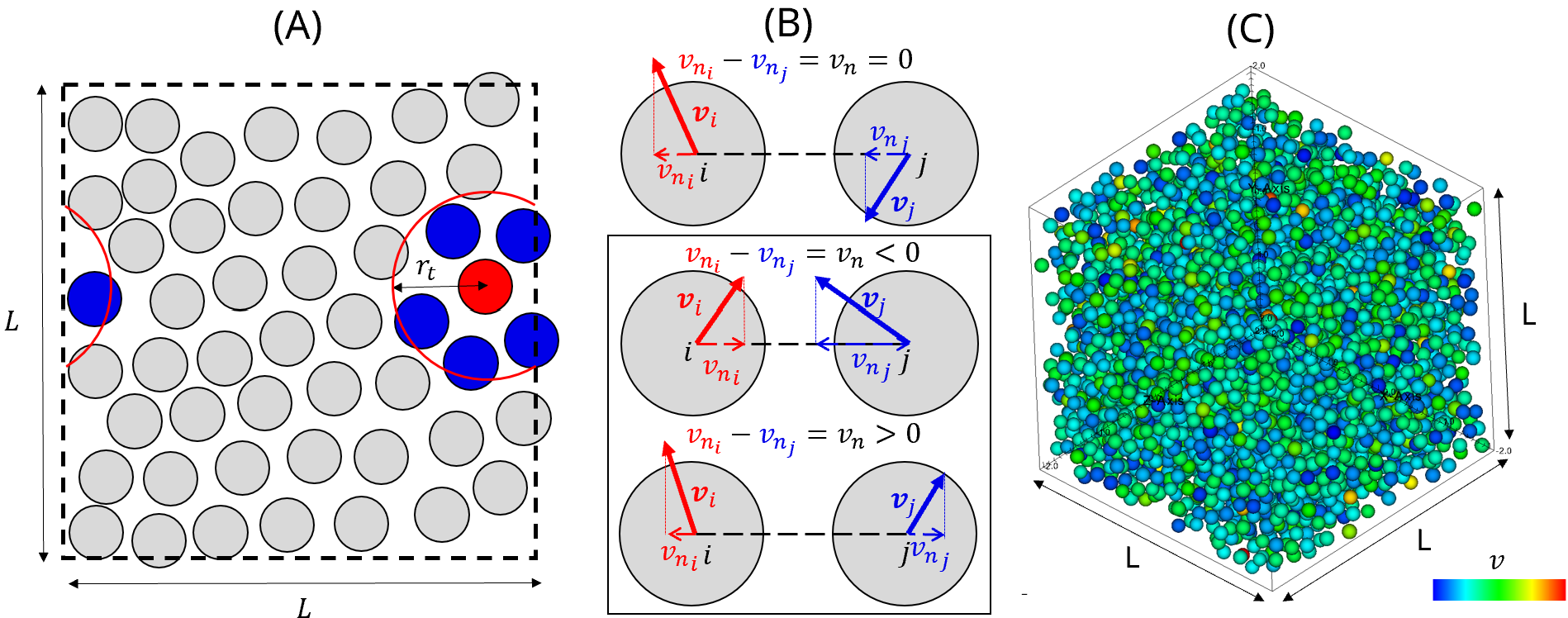}
    \caption{\label{fg_pairwise_example}(A) Schematic representation of the constant volume periodic cell. Given a particle (red), its thermostat pairs (blue) are identified within the distance $r_t$. (B) Relative velocities of a particle pair projected onto the normal vector. Outside the box, the particles have coherent motion (zero thermal energy, $v_n=0$). (C) Screenshot of the granular medium during a simulation.}
\end{figure*}

Both the DEM code and the thermostats are implemented within the open-source library Mechsys (mechsys.nongnu.org, see \citep{galindo2013coupled}). The code is structured to parallelise most calculations for GPU acceleration through the CUDA API. For example, race conditions are avoided by using the \textit{atomicAdd} function, which performs atomic addition to global variables from each GPU thread. For simplicity, the DEM cycle is summarised into four main steps: (i) contact detection; (ii) force calculation; (iii) particle position/velocity update; and (iv) force reset. Contact detection is the most computationally expensive portion of a DEM cycle, and every DEM code adopts broad-phase filtering techniques to restrict detailed contact analysis to a small portion of contact candidates. Mechsys employs a Verlet lists approach for this task: for each particle, a list is constructed that identifies all possible contact candidates within a cut-off threshold (the Verlet distance), which is then used for the successive contact analysis. When two particles move further apart than this distance, the list is updated. Mechsys performs this first step sequentially (i.e. on the CPU), before transferring the data to the GPU to parallelise contact resolution. To improve efficiency, the code employs a large Verlet distance (e.g. $\ge2d$) and static list length. This design reduces the frequency of Verlet list calculations, and therefore the data transfer between CPU and GPU, which is the main computational bottleneck. Although a large number of potential collision pair candidates are discarded during the contact detection step, the static list length allows contact resolution (validating the contact and calculating the force to add to each particle) to be carried out on individual GPU threads, taking advantage of multi-threaded parallelisation. \\
\indent The thermostat equations are implemented following the pairwise DPD approach, in order to retain Galilean invariance and conserve pair centre-of-mass momentum. Because the pairwise approach follows the same structure of a generic contact interaction calculation, the thermostat forces are also computed within the force calculation step, repurposing the existing data structure and adding minimal computational overhead. Given a pair of particles $i,j$ at position $\boldsymbol{x}_{i,j}$, the thermostat is applied if $||\boldsymbol{x}_i-\boldsymbol{x}_{j}||<r_t$, where the $||\cdot||$ is the norm and $r_t$ is the interaction range $r_t$ (Figure \ref{fg_pairwise_example}A). A smoothing function is used to minimise the influence of particles at the edge of the interaction range, reducing numerical noise and sensitivity to rounding errors \citep{allen_thermostat_2006,phan2013understanding,verbeek_advantages_2022}. Herein, we adopt a smooth Heaviside function, $H=0.5(1+\tanh[(r_t-||\boldsymbol{x}_i-\boldsymbol{x}_{j}||)/r_h])$, where $H$ is the correction factor [0,1] applied to the force and $r_h$ is the width of the interpolation distance. 
Given particle pair radii $r_i,r_j$, their average $r_e=(r_i+r_j)/2$ is used to define the dimensionless parameters $R_t=r_t/r_e$ and $R_h=r_h/r_e$. The relative particle velocity $\boldsymbol{v}_r=\boldsymbol{v}_i-\boldsymbol{v}_j$ is projected onto the normal vector $\boldsymbol{n}=(\boldsymbol{x}_i-\boldsymbol{x}_{j})/||\boldsymbol{x}_i-\boldsymbol{x}_{j}||$ to obtain the projected relative velocity $v_n$ (Figure \ref{fg_pairwise_example}B). 
The pairwise Nos\'e-Hoover formulation proposed by \cite{allen_thermostat_2006} is
\begin{eqnarray}\label{eq_nose-hoover_pt1}
\boldsymbol{F}_\text{NH}=\boldsymbol{n}m_rv_n\zeta,
\end{eqnarray}
where $m_r=m_im_j/(m_i+m_j)$ is the reduced mass of the interaction. The $\Delta\zeta$ increment over the time step $\Delta t$ is based on the average difference between the pairwise temperature ($T_p$) and the prescribed temperature ($\hat{T}$); as given by
\begin{equation}\label{eq_nose-hoover_pt2}
\Delta{\zeta}=(\Delta t/Q)(\langle{T_p-\hat{T}}\rangle/\hat{T}).
\end{equation}
 The pairwise temperature is defined as $T_p=v_n^2/2$ because the 1D projection splits the magnitude of the velocity fluctuation equally between the two particles. After some tentative simulations, we decided not to adopt a chain implementation because it did not seem to provide any improvement to system stability. 
 Therefore, in this single-thermostat implementation, eq. (\ref{eq_nose-hoover_pt1}) acts as a servo-control with stiffness $\Delta t/Q$, steering the temperature of the system towards the target. In this framework, an overly \textit{stiff} thermostat introduces a systematic bias: eq. (\ref{eq_nose-hoover_pt2}) prescribes a minimum increment of $-\Delta t/Q$ and a maximum of infinity. If the system is unstable and it is undergoing oscillations with a magnitude greater than the prescribed temperature ($|\langle\Delta T_p\rangle|>\hat{T}$), the asymmetry in the minimum and maximum $\Delta \zeta$ values produces a non-zero (positive) mean (i.e. cooling). This results in a temperature close to zero for most of the simulation steps with a few positive spikes bringing the time-average back to the prescribed value. Such a system is highly non-ergodic, producing distorted hydrodynamics properties. While this limitation is at the core of the Nos\'e-Hoover approach, it does not represent a major concern in MD due to a combination of low timesteps, chain implementation, smooth fields (e.g. velocity) and interactions (e.g. Lennard-Jones potential). Here it can become an issue for pure Nos\'e-Hoover simulations, because the velocity-segregated system that arises from unconstrained inelastic collisions is prone to numerical instabilities. \\
 \indent In the discrete implementation, the Wiener increment is $dW=R\sqrt{\Delta t}$, with being a random value $R$ independently sampled from a standard normal distribution at each simulation step. The adopted pairwise Langevin formulation is
\begin{equation}\label{eq_langevin}
\boldsymbol{F}_\text{La}=\boldsymbol{n}m_e\left(-v_n\xi_\text{La}+R\sqrt{2\xi_\text{La}\hat{T}/\Delta t}\right),
\end{equation}
where the equivalent mass $m_e$ is twice the reduced mass ($m_r$). The stochastic term of the Scaled Langevin formulation is
\begin{equation}\label{eq_simplified_langevin}
\boldsymbol{F}_\text{LaS}=\left\{ 
\begin{aligned}
    &\boldsymbol{n}m_eR\sqrt{2\xi_\text{La}(\hat{T}-T_p)/\Delta t} &\quad\text{if}\  2\hat{T}-v_n^2>0,\\
    &\boldsymbol{0}&\quad\text{if}\ 2\hat{T}-v_n^2\leq0.
\end{aligned}   
\right \}
\end{equation}
Due to its pairwise implementation, the stochastic amplitude is largest for low-relative-velocity pairs (\ref{eq_simplified_langevin})  and decreases as their pairwise kinetic state approaches the target temperature.\\
\indent An individual pseudo-random number generator instance (cuRAND State) is assigned to each thread to solve eq. (\ref{eq_langevin}) and eq. (\ref{eq_simplified_langevin}). At the end of the force calculation step, $\Delta T_p=v_n^2-2\hat{T}$ and $H$ are summed and passed to a single thread through block reduction and warp shuffling, minimising the number of atomicAdd calls. Particle motion is then solved using the standard Verlet algorithm and time is updated ($t:=t+\Delta t$). Finally, the global Nos\'e-Hoover friction parameter $\zeta$ is calculated on that single GPU thread during the force-reset portion of the DEM cycle. Equation (\ref{eq_nose-hoover_pt2}) is solved using explicit Euler integration, with $\langle T_p\rangle=\sum\Delta T_p/\sum H$ (i.e. distance-weighted average with Heaviside thresholding).
 \subsection{Simulation programme}\label{sc_simulation_programme}
The performance of the different thermostat formulations is evaluated through a set of parametric studies.
First, the combined effect of interaction distance ($R_t$) and coupling strength ($\xi_\text{NH,La}$) is investigated at constant target temperature ($\hat{T}=10^{-1.5}$, units of $L^2T^{-2}$), timestep size ($\Delta t=10^{-6}$, units of $T$), and DEM properties (Table \ref{tb_dem_parameters}). For pure Nos\'e-Hoover and Langevin simulations, the following values are used: $\xi_\text{La}$ = [$10^2,10^{2.5},10^3$] (units of $T^{-1}$) and $\xi_\text{NH}$ = [$10^{0}$, $10^{-1}$, $10^{-2}$] (units of $T$). For the hybrid formulations, we employ the same Nos\'e-Hoover coupling strength values used in the fully deterministic scenario, while the strength of Langevin coupling is lowered, because it no longer has to enforce temperature control: $\xi_\text{La}$ = [$10^{-1}$, $10^{0}$, $10^{1}$, $10^{2}$], for a total of 12 combinations. For each of these, six interaction ranges are tested, $R_t$ = [2.0, 2.5, 2.75, 3.0, 3.5, 4.0] (non-dimensional), for a total of 180 simulations.\\
\indent \cite{allen_thermostat_2006} notes that $\Delta t$ can affect the thermostat response. After some tentative simulations, we observed that pure Nos\'e-Hoover became unstable and prone to crashes at $\Delta t>10^{-6}$ (depending on both thermostat and DEM parameters); while the results obtained under pure Langevin exhibited timestep-dependence at $\Delta t\ge10^{-5}$. On the other hand, the hybrid formulations appeared to retain stability and timestep independence. Therefore, we sought to determine whether they would maintain this desirable property as $\Delta t$ approaches the critical time step for numerical stability (e.g. $2\sqrt{m/K_n}\approx5\cdot10^{-4}$). The second set of simulations therefore concerns time step size at different levels of kinetic energy (i.e. $\hat{T}$).  
We test five temperatures, $\hat{T}$ = [$10^{-3},10^{-2.5},10^{-2},10^{-1.5},10^{-1}$] and six timestep sizes, $\Delta t$ = [$10^{-6}$, $5\cdot10^{-6}$, $1\cdot10^{-5}$, $2.5\cdot10^{-5}$, $5\cdot10^{-5}$, $10^{-4}$].\\
\indent Dimensional analysis is used to obtain a consistent thermostat response. Since the coupling parameters of the thermostat ($\xi_\text{La}$ and $\xi_\text{NH}$) have units of inverse time and time, their product defines the dimensionless quantity $\pi_1=\xi_\text{La}\xi_\text{NH}$, which parametrises the relative response scales of the stochastic and deterministic components, similar to the approach by \cite{stoyanov_molecular_2005}. While section \ref{sc_nosehoover_Pv} shows that $\xi_\text{NH}$ has no influence in a homogeneous steady state system, the simulation will inevitably include some degree of inhomogeneity and oscillations, which would lead to this parameter exhibiting an influence on the dynamics and transport properties \citep{garzo2002transport,j2007statistical}. The stochastic coupling is characterised through
 \begin{equation}\label{eq_T_nh}
     \pi_2=\sqrt{\hat{T}}/(\xi_\text{La}d),
 \end{equation}
where $d/\sqrt{\hat{T}}$ is the characteristic particle-motion timescale and $\xi_\text{La}^{-1}$ is the stochastic coupling timescale. Therefore, smaller values of $\pi_2$ correspond to faster, stronger stochastic decorrelation relative to the particle dynamics.  
These two numbers are incorporated in the second parametric study: $\pi_1=0.01$ is kept constant, and three values of $\pi_2$ = [$0.71,1.66,2.24$] are tested. Given the two hybrid formulations (NA and SL), this results in a total of 180 simulations.\\
\indent The third parametric study concerns the degree of inelasticity ($e$ = [$1.0, 0.9, 0.75,0.5,0.4,0.2,0.1$], non-dimensional). The coupling parameters are kept constant ($\xi_\text{La}=1.0$, $\xi_\text{NH}=0.01$), with different temperatures (same as study \#2), resulting in five $\pi_2$ values ($0.16,0.28,0.5,0.89,1.58$). Timestep size is tested again, since the degree of inelasticity affects the development of inhomogeneous conditions \citep{garzo2004diffusion,kang2010granular} and could therefore produce instability. Still, since it is not the main focus of this study, we only test four timestep sizes ($\Delta t$ = [$10^{-6}$, $4.64\cdot10^{-6}$, $2.15\cdot10^{-5}$, $10^{-4}$]), for a total of 280 simulations. 
The three parametric studies are summarised in Table \ref{tb_param_study}.
\begin{table}[h]
\caption{\label{tb_param_study} List of parametric studies. Unless otherwise specified, all DEM parameters are set according to Table \ref{tb_dem_parameters}.}
\begin{ruledtabular}
\begin{tabular}{cccr}
\textrm{Study \#}&
\textrm{Varying}&
\textrm{Constant}&
\textrm{Thermostats}\\
\colrule
   1 & $R_t,\ \xi_\text{La},\ \xi_\text{NH}$& $\hat{T},\ \Delta t$ & LA, NH, NA, SL\\
   2 & $\Delta t,\ \pi_2,\ \hat{T}$& $\pi_1,\ R_t$& NA, SL\\
   3 & $\Delta t,\ e,\ \hat{T}$& $\xi_\text{La},\ \xi_\text{NH},\ R_t$& NA, SL 
\end{tabular}
\end{ruledtabular}
\end{table}
\subsection{Data processing}\label{sc_data_processing}
The simulation data is processed to quantify two distinct aspects of thermostat: (i) its ability to enforce the prescribed scalar granular temperature, and (ii) its influence on the statistical state and emerging observables. While both the thermostat parameters and the DEM properties influence the time required to reach stationarity, we assume that all simulations presented here are under steady‑state conditions after 50\% of the total simulation time $t_f$, and the data sampled within this interval is thereby used for postprocessing.\\
\indent Obtaining a unique ensemble-average temperature presents the same difficulties encountered in the definition of a Galilean invariant thermostat: measuring the temperature of an individual particle requires the definition of a local group velocity, which itself depends on the definition of a characteristic length scale. Since granular materials do not possess a strong length scale separation \citep{goldhirsch2008introduction}, the choice of this parameter is largely arbitrary. The pairwise inertial frame of reference approach used to enforce temperature control introduces a systematic sampling bias: dense clusters are oversampled, while isolated particles are ignored. Because the latter are statistically more likely to represent high-velocity fluctuations \citep{goldhirsch2008introduction}, the pairwise average underestimates the true temperature. Moreover, if post-processing uses the same $R_t$ as the thermostat, it only samples the portion of the system where temperature control is enforced, which will always appear closer to the target value than the true domain average. This characteristic, however, makes $\langle T_p\rangle$ useful to assess thermostat performance. For example, the Langevin thermostat is unable to enforce thermalisation in a dissipative system, as shown in eq. (\ref{eq_temperature_scaling_langevin}). The local ratio between measured and target temperature, $\langle {T_p}\rangle/\hat{T}$, therefore serves as a proxy for the average value of $\Gamma$.\\
\indent If the stochastic term successfully mitigates the formation of cold clusters, particle velocities become spatially uncorrelated, and the group velocity $\boldsymbol{v}_g$ is zero (since the inertial frame of reference of the system is at rest). Under such conditions, $\langle T\rangle$ is a suitable approximation for the true temperature. Conversely, a large discrepancy in which $\langle {T}\rangle\gg\hat{T}$ while $\langle{T_p}\rangle\simeq\hat{T}$ indicates coherent particle motion, consistent with velocity segregation or flying-ice-cube behaviour \citep{braun2018anomalous}. If $\langle {T}_p\rangle \gg \langle{T}\rangle$, the thermostat is enforcing the temperature only on a subset of particles that is too small for thermal diffusion to homogenise the system, indicating that the interaction range $R_t$ is too low. \\
\indent  Thermostat accuracy is quantified using the time-averaged temperatures $\langle\bar{T}\rangle$ and $\langle\bar{T_p}\rangle$. For statistical significance, the pairwise temperature $\langle T_p\rangle$ is preferred over $\langle T\rangle$, since the former is a thermostat variable and it is computed at runtime, while the latter is obtained in postprocessing from a smaller number of output steps. When spatial sampling biases are present (i.e. $R_t<3$), both variables are reported. Additional quantities of interest are the velocity kurtosis $\bar{\kappa}(v)$ and the 5th and 95th velocity percentiles ($P_{5}$, $P_{95}$), which capture anomalies in the velocity distribution; as well as the standard deviation of the local temperature error over time, $\sigma(\langle \Delta T_p\rangle)$, where $\langle \Delta T_p\rangle=\langle  T_p\rangle-\hat{T}$, which quantifies thermostat precision. All velocity variables are normalised using $\hat{T}$.\\
\indent To quantify the mesoscopic consequences of the different energy-transfer mechanisms, we employ standard metrics such as velocity autocorrelation and mass diffusivity to assess the meso-scale properties of the system \citep{pastorino2007comparison,qian2009effective,basconi_effects_2013,leimkuhler2015numerical}. Linking these variables to measurable physical quantities relies on assumptions (e.g. homogeneity) that are not necessarily valid for dissipative systems \citep{garzo2002transport,gomart2004granular,shah2025molecular}. Therefore, the resulting values should only be used for quantitative comparison between simulations and not be interpreted as true measures of emergent material properties. In particular, it was not possible to use the data to quantify linear viscosity, showing that, despite the contact-chain-breaking influence of the thermostat, non-Markovian effects are significant enough to prevent the application of this framework \citep{campbell2002granular}, and it is therefore included as a \textit{failed result} in section \ref{sc_viscosity}.\\
\indent The temporal persistence of particle motion is quantified through the Velocity AutoCorrelation Function, VACF$(\tau)=\langle\overline{\boldsymbol{v}(t+\tau)\cdot\boldsymbol{v}(t)}\rangle/3$, where $t$ indicates an arbitrary starting time, $\tau$ is the time lag at which the correlation is measured and the large bar indicates that the value is time-averaged (over the different starting times). The rate of decay of this function provides information on how the different thermostats make the system dynamics decorrelate over time (e.g. through fictitious heat bath collisions) \citep{ruiz2018effect}. Diffusivity can then be quantified from the VACF through the Green-Kubo relation, although it is often more practical to measure this property by fitting a linear relationship on the Mean Square Displacement (MSD) of the particles
 \begin{equation}\label{eq_MSD}
     D=\lim_{\tau\rightarrow\infty}\overline{\langle||\boldsymbol{r}(t+\tau)-\boldsymbol{r}(t)||^2\rangle}/6\tau,
 \end{equation}
 where $\boldsymbol{r}(t)=\boldsymbol{x}(t)-\boldsymbol{x}(t_0)$ is the displacement of an individual particle at time $t$. Due to the periodic boundary conditions, minimum image convention is used to correct $\boldsymbol{x}$, assuming that any given particle can cross the periodic boundary at most once in a given time increment, i.e. $||\boldsymbol{x}(t+\Delta t)-\boldsymbol{x}(t)||<L$.
  The Radial Distribution Function (RDF) describes the spatial distribution of the particles in the domain, normalised by the domain average ($N/V$, with $N$ being the total number of particles in the domain). A rarefied (ideal) gas with no particle interactions has no excluded volume effects and it is therefore characterised by a uniform value of $g(r)$, resulting in a flat line ($g\approx1$), while local peaks and troughs indicate preferential relative positions, i.e. a lattice structure which locks particles in place, inhibiting diffusion. The particle density $g$ at the radius $r$ is obtained as
 \begin{equation}\label{eq_rdf}
     g(r)=\frac{V}{4\pi r^2 N^2}\overline{\sum_{i=1}^N\sum_{i\neq j}\delta \bigl(||\boldsymbol{x}_i-\boldsymbol{x}_j||-r\bigl)},
 \end{equation}
 where $\delta$ is the Dirac delta function. In practice, $\delta$ is replaced by discrete bins: particles are placed within spherical shells of thickness $\Delta r$ and eq. (\ref{eq_rdf}) becomes $g(r)=V\overline{\sum\sum1(r\leq||\boldsymbol{x}_i-\boldsymbol{x}_j||<r+\Delta r)}/(4\pi r^2\Delta rN^2)$, where 1 is the indicator function. As with the MSD calculation, the minimum image convention is applied to correct distances across the periodic boundaries. \\
 \indent For dense systems where excluded volume effects are significant, the radial distribution function peaks at the particle contact distance ($r/d\approx1$). The associated value, $g(d)$, enters kinetic-theory expressions for the mean free path, collision frequency and transport properties \citep{garzo2002transport,berzi2024granular}. 
 In this context, \cite{carnahan1969equation} provides an analytical approximation for a system of non-attractive, monodisperse hard spheres
\begin{equation}\label{eq_carnahan_approx}
g(d)=(1-0.5\phi)/(1-\phi)^3\approx2.22.
\end{equation}
This value was initially adopted as a reference to quantify the increase in clustering due to particle inelasticity \citep{goldhirsch2008introduction}. However, numerical simulations with fully elastic spheres show that the slightly polydisperse nature of the system presented here causes a reduction in $g(d)$ \citep{ogarko2012equation}. Therefore, we adopt $g(d)=2$ as the reference value for this study, rather than the one from the Carnahan-Starling approximation.
\section{Results and discussion} \label{sc_results}
\subsection{Individual thermostat comparison}\label{sc_individual_comparison}
\begin{figure}[!htb]
    \centering
    \includegraphics[width=\linewidth]{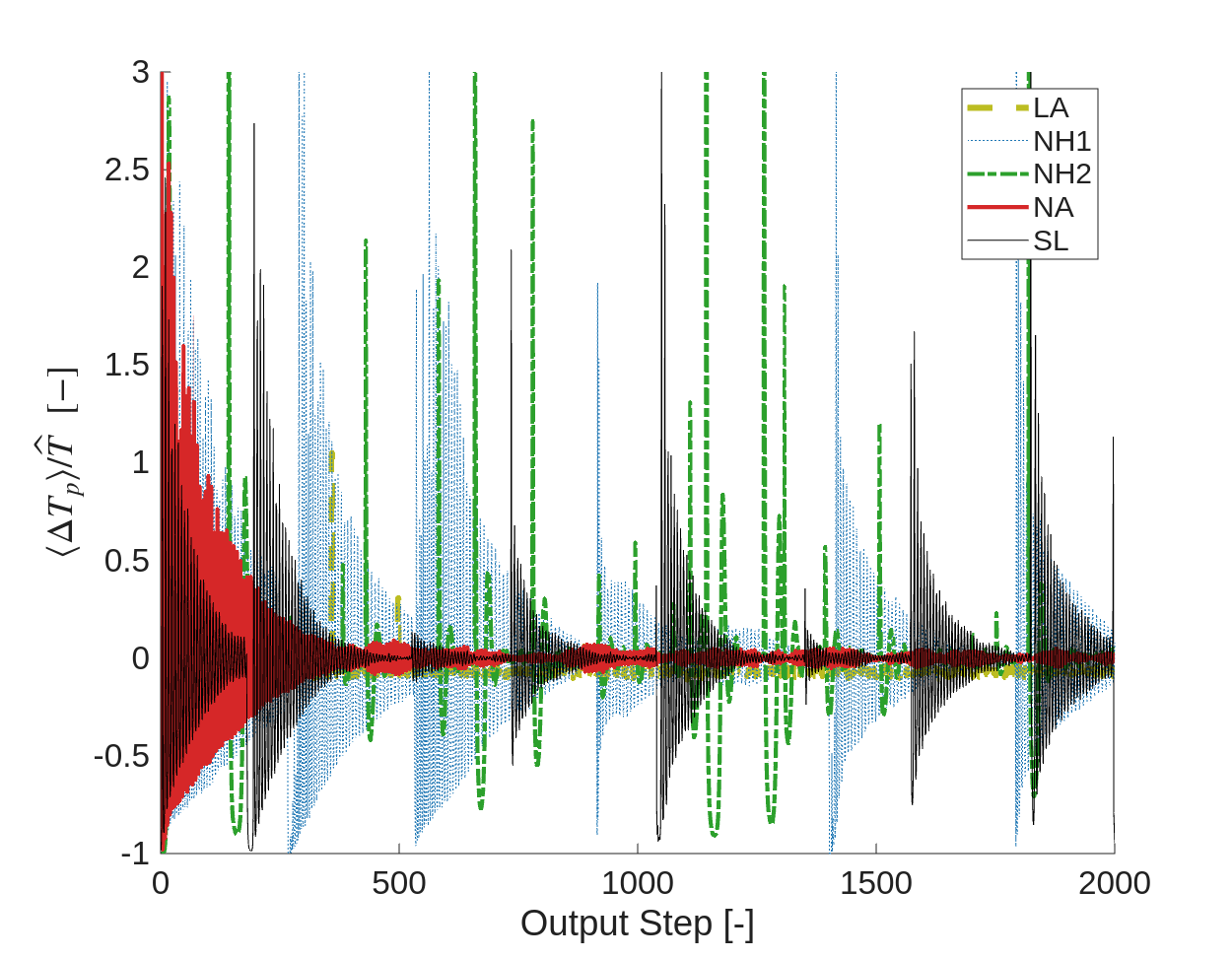}
    \caption{Example of the temperature oscillations observed for a fixed $R_t=3$ and different thermostats for the first 2000 output steps: LA, $\xi_\text{La}=10^2$; NH1, $\xi_\text{NH}=10^{-2}$; NH2, $\xi_\text{NH}=10^0$. NA and SL adopt the same parameters ($\xi_\text{La}=10^{-1},\ \xi_\text{NH}=10^{-2}$). Note that the maximum value of $\langle\Delta T_p\rangle/\hat{T}$ is capped at 3 in the plot, while some of the Nos\'e-Hoover induced oscillations go up to 14.}
    \label{fg_granrand_Time_Temperature_pair}
\end{figure}

\begin{figure}[!htb]
    \centering
    \includegraphics[width=\linewidth]{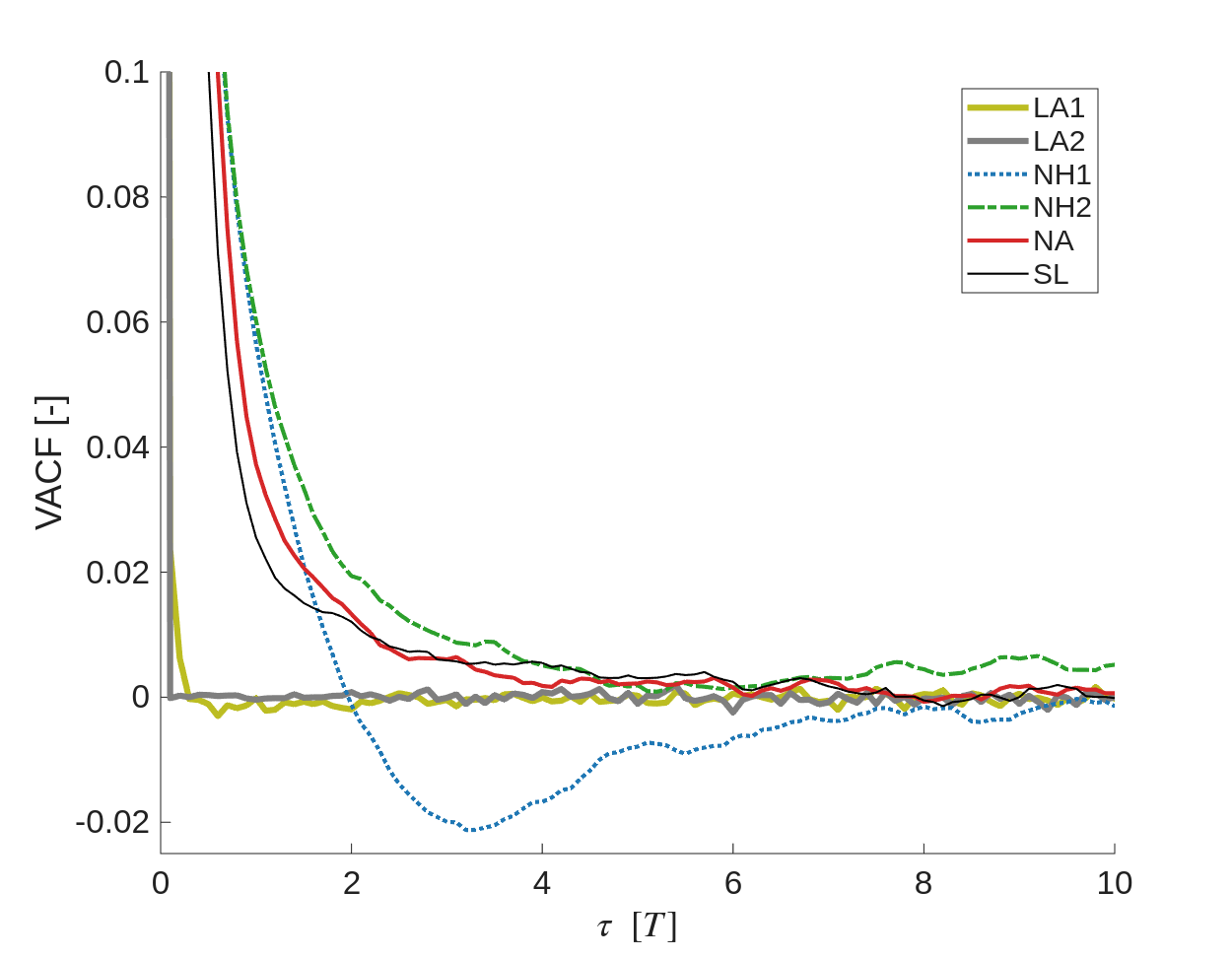}
    \caption{Example of the VACF plot for the different thermostats. LA1, $\xi_\text{La}=10^2,\ R_t=3$; LA2, $\xi_\text{La}=10^3$, $R_t=4$. NH1, $\xi_\text{NH}=10^{-2}$; NH2 $\xi_\text{NH}=10^0$. NA and SL have the same parameters ($\xi_\text{La}=10^0,\ \xi_\text{NH}=10^{-2}$). $R_t$ is set to 3 in all scenarios, except for (LA2) where it is set to 4 to highlight the stronger effect of Langevin dynamics at increased $R_t$ \citep{verbeek_advantages_2022}.}
    \label{fg_VACF_example}
\end{figure}

\begin{figure}[!htb]
    \centering
    \includegraphics[width=\linewidth]{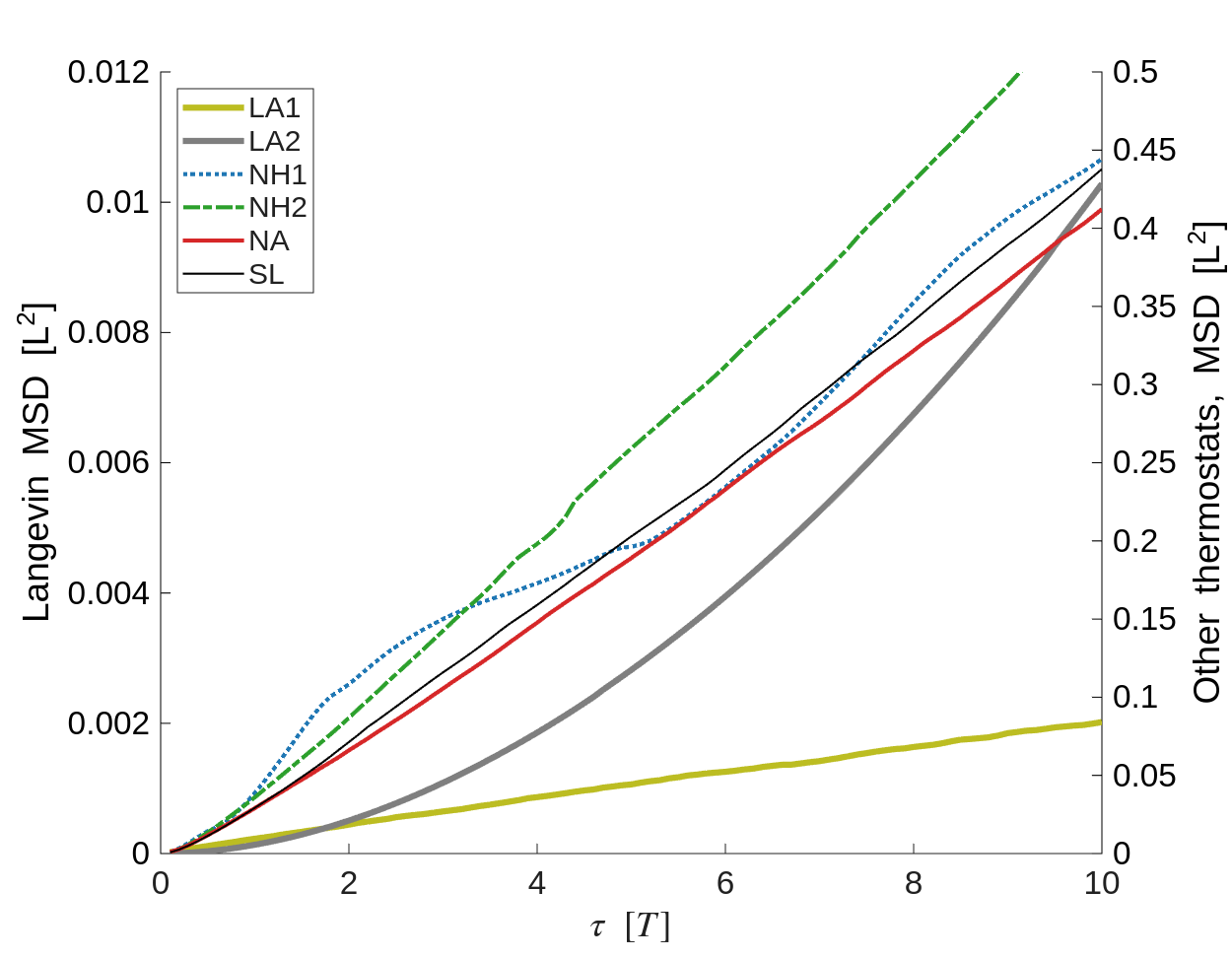}
    \caption{Example of the MSD evolution for the different thermostats. The same legend as Figure \ref{fg_VACF_example} is used here.}
    \label{fg_MSD_lag}
\end{figure}

\begin{figure}[!htb]
    \centering
    \includegraphics[width=\linewidth]{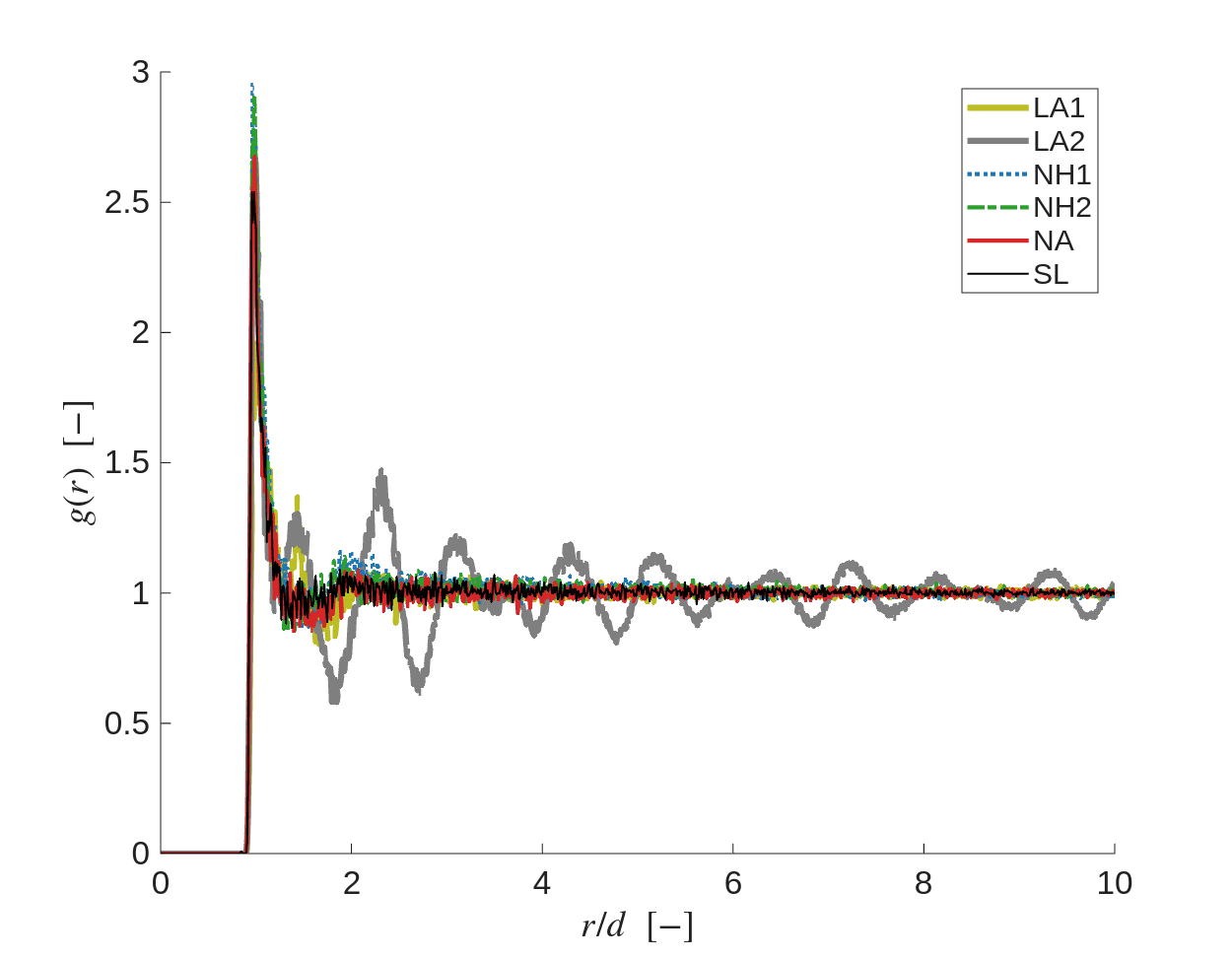}
    \caption{Example of the RDF distribution for the different thermostats. The same legend as Figure \ref{fg_VACF_example} is used here.}
    \label{fg_RDF_example}
\end{figure}

In the following, the different thermostats are qualitatively compared using data obtained from the first parametric study (constant temperature).\\
\indent Figure \ref{fg_granrand_Time_Temperature_pair} shows the time-evolution of temperatures. This is done to illustrate the behaviour of the different thermostats and to explain the trends observed in section \ref{sc_parametric_study}, not to demonstrate whether one formulation is superior to the others. Since all the data is obtained with $R_t=3$, $\langle \Delta T_p\rangle$ is used as the representative variable. Results from pure Langevin and pure Nos\'e-Hoover simulations are respectively indicated as LA and NH, while the Naive Addition of the thermostats is indicated as NA and the Scaled Langevin approach as SL. \\
\indent First of all, as demonstrated in eq. (\ref{eq_temperature_scaling_langevin}), pure Langevin is unable to enforce the target temperature in a dissipative system: the depicted coupling strength ($\xi_\text{La}=100$) results in a consistent $\approx10\%$ error, $\langle\Delta\bar{T}_p\rangle/\hat{T}\approx-0.1$. Further increasing $\xi_\text{La}$ (up to 1000) and the interaction range (to $R_t=4$) only brings the error down to $\approx5\%$. As expected, the time-averaged temperature matches the target for both pure Nos\'e-Hoover thermostats, regardless of the $\xi_\text{NH}$ value. Unfortunately, these simulations are also characterised by large temperature spikes (local $\langle T_p\rangle\gg\hat{T}$). These fluctuations are not caused by the excessive thermostat stiffness (i.e. $|\langle\Delta T_p\rangle/\hat{T}|\ll1$), but by the underlying highly non-ergodic conditions (velocity segregation), and occur even when the system is approaching steady state $\Delta\zeta\rightarrow0$. In fact, both Nos\'e-Hoover scenarios have the same peak oscillation magnitude ($\langle T_p\rangle\ge 14\hat{T}$), despite possessing vastly different thermostat stiffnesses: given $Q=\xi_\text{NH}^2$, NH1 ($\xi_\text{NH}=0.01$) is four orders of magnitude more stiff than NH2 ($\xi_\text{NH}=1.0$). The coupling parameter however does affect the post-spike response, with NH1 behaving as a dampened oscillator (asymmetrical and centred in $\langle \Delta T_p\rangle=0$), and NH2 returning the temperature to the target within a single (slow) oscillation. Regarding the hybrid formulations, the introduction of stochastic kicks disrupts the correlated states that develop under pure Nos\'e-Hoover, thereby improving numerical stability. The specific scenario depicted ($\xi_\text{La}=0.1$) is chosen because it highlights that the Naive Addition (NA) formulation prevents temperature spikes even when the stochastic term is small, while SL does not. At the same time, SL results are more accurate (in terms of $\langle\Delta T_p\rangle$ being closer to zero), as can be observed from the oscillation magnitudes at the beginning of the simulation and away from the $\langle\Delta T_p\rangle$ spikes (e.g. around step 1000).\\ 
\indent Moving onto the effect of the thermostat on the dynamics of the system, Figure \ref{fg_VACF_example} shows the loss of velocity autocorrelation (VACF) for a few selected simulations. As reported in the literature, e.g. \cite{basconi_effects_2013}, Langevin significantly dampens the underlying system dynamics (VACF$\rightarrow0$ at $\tau\le0.2$), even when the thermostat action is insufficient to enforce the target temperature (LA1, $\langle\Delta\bar{T}_p\rangle/\hat{T}\approx-0.1$). When the coupling strength is increased (LA2), the correlation goes to zero within a single lag increment ($\tau= 0.1$). The stiffness of the deterministic term indirectly affects the hydrodynamic behaviour through its response to temperature spikes. Although NH1 and NH2 initially ($\tau\le1$) show the same VACF decay, the former behaves as a dampened oscillator, leading to negative VACF values: the particle velocities become negatively correlated at regular time lags, and the correlation returns to zero as the oscillations decay as is typical of coherent oscillators \citep{hoover2004time}. The VACF plot for NH2 does not develop any negative correlations because the temperature oscillations in this scenario are poorly correlated in time, and particle velocities do not become fully uncorrelated within $\tau\le10$. Regarding the hybrid formulations, they both exhibit the same long-term behaviour, but differ within $\tau<2.5$, with SL consistently decorrelating faster than NA. This indicates that, at least for weak stochastic coupling (e.g. $\xi_\text{La}=1.0$), the damping term $-v_n\xi_\text{La}$ is not the most significant disruptor of system dynamics. We hypothesise that the rate of autocorrelation loss differs between SL and NA because the stochastic forcing in SL is state dependent: its amplitude is maximum for fully correlated pairs ($T_p=0$) and progressively decreases as $T_p$ approaches $\hat{T}$, whereas the stochastic amplitude in NA is independent of the instantaneous pair kinetic state.\\ \indent Figure \ref{fg_MSD_lag} shows the evolution of the mean squared displacements (MSD), which quantify diffusion. Fully stochastic temperature control (Langevin) is characterised by such a negligible diffusivity value that its MSD-$\tau$ evolution has to be plotted on a different y axis (left) than the other thermostats (right y axis). If an ideal MSD plot exhibits an initial power law evolution (ballistic regime), followed by a linear trend (Fickian diffusivity), Langevin dynamics do not leave the first stage. The power-law evolution can be made apparent by lowering the number of particles the thermostat is applied to ($R_t=2.0$, LA2), while an increase in the coupling strength and range ($R_t\ge3$, LA1) produces an over-dampened scenario, which appears linear within the time-scale of the simulation. Since the actual displacement values are negligible (MSD$/d^2\le0.05$ at $\tau=10$), it is assumed that the linear trend is not indicative of Fickian diffusion. 
Moving to Nos\'e-Hoover results, the scenario that behaved like a dampened oscillator in Figure \ref{fg_granrand_Time_Temperature_pair} (NH1, $\xi_\text{NH}=0.01$), and showed negative VACF, here exhibits a sinusoidal behaviour on top of the linear trend. This is again attributed to the high regularity of the post-spike thermostat response \citep{hoover2004time}. In contrast, NH2 ($\xi_\text{NH}=1$), exhibits oscillations of similar magnitude but lacks the same high degree of correlation. Consequently, it does not produce the sinusoidal pattern and it is substantially more diffusive, matching the trends of the Figure \ref{fg_VACF_example}. Both hybrid formulations (NA and SL) exhibit Fickian behaviour, with SL exhibiting slightly higher diffusion (which contrasts with the slightly higher rate of correlation loss shown in Figure \ref{fg_VACF_example}). For the successive comparison (section \ref{sc_parametric_study}), diffusivity is quantified by fitting a line  to the MSD$-\tau$ curve (for $\tau>2.5$). We do not use the rate of velocity-correlation loss as an additional comparative metric because its long-term information is largely reflected in the MSD-derived diffusivity, while VACF is more sensitive to temporal discretisation.\\
\indent The Radial Distribution Function (RDF) plot (Figure \ref{fg_RDF_example}) can be used to explain the low diffusivity values. The degree of clustering in the system is quantified by the peak particle density observed at contact ($r/d\approx1$, with $d$ being the reference particle diameter). 
Since the Nos\'e-Hoover formulation does not counteract the clustering effect of inelastic collisions, it is characterised by the highest peak value, i.e. $g(d)\approx2.9$. On the opposite end of the spectrum, the Langevin thermostat (scenarios LA1 and LA2) matches the peak obtained for fully elastic spheres, i.e. $g(d)\approx 2$, but it also introduces a secondary peak at $r/d=\sqrt{2}$, indicating crystallisation (preferential placement in a hexagonal packing structure) \citep{bai2019crystallization}. The scenarios previously identified as over-dampened Langevin dynamics ($R_t=4$ in Figures \ref{fg_VACF_example} and \ref{fg_MSD_lag}) further extend the crystallisation pattern (LA2, $\xi_\text{La}=1000$, $R_t=4$). The plot exhibits a clear sinusoidal distribution, which indicates that the crystallisation (preferential relative particle positions) is not limited to isolated clusters (e.g. within $r<2$), but that the lattice structure spans across a large portion of the domain. Under these conditions, particles vibrate at the target temperature without diffusing (significant temporal and spatial variations would flatten the $g(r)$ distribution) \citep{ruiz2018effect}.
\subsection{Parametric studies}\label{sc_parametric_study}
\subsubsection{Study \#1, changing $\xi_\text{La}$, $\xi_\text{NH}$ and $R_t$ at constant $\hat{T},\Delta t$}
\begin{figure*}[!htb]
    \centering
    \includegraphics[width=1.0\linewidth]{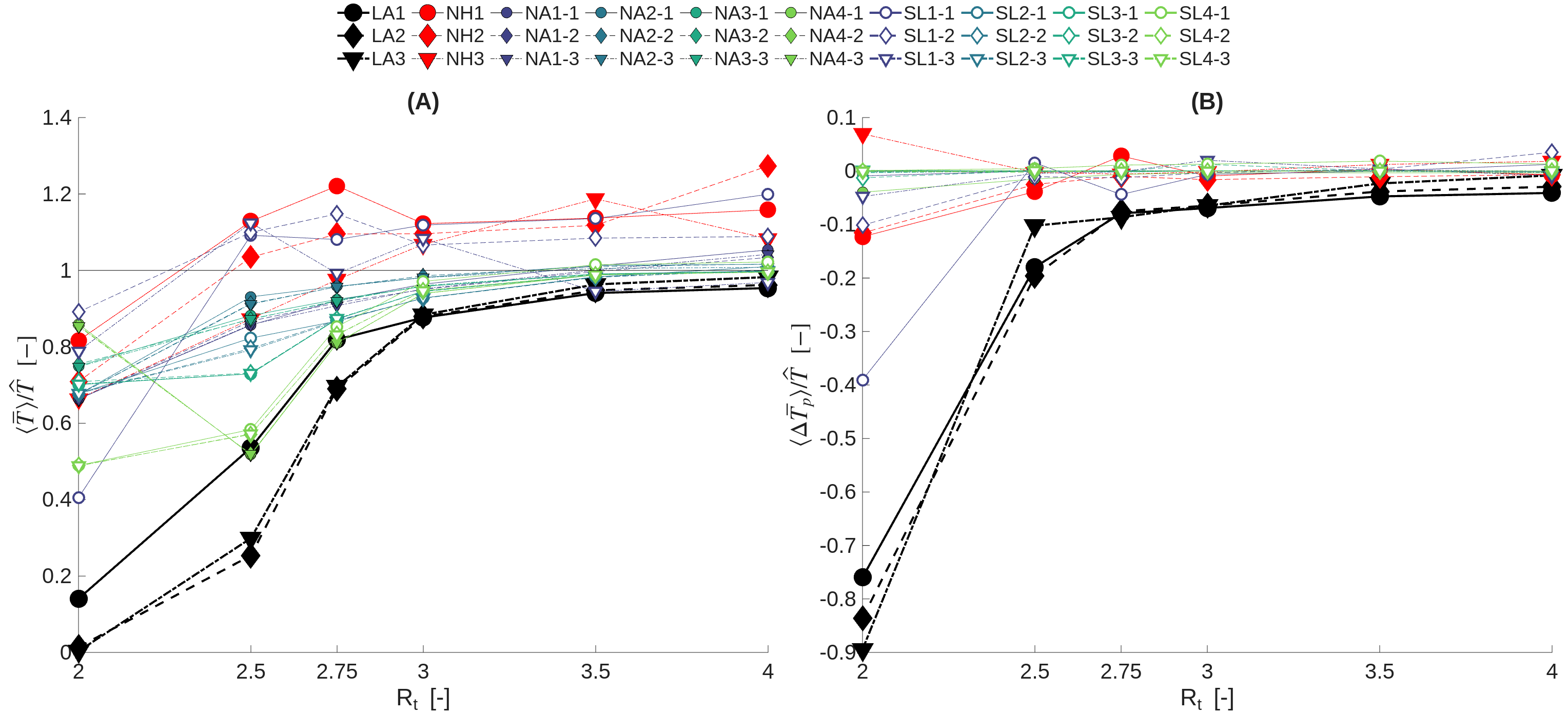}
    \caption{Effect of $R_t$ and thermostat coupling parameters ($\xi_\text{La}$ and $\xi_\text{NH}$) on temperature control. Subplot (A) indicates $\langle \bar{T}\rangle/\hat{T}$, subplot (B) $\langle \Delta\bar{T_p}\rangle/\hat{T}$. }
    \label{fg_granrand_T_Rt}
\end{figure*}

\begin{figure*}[!htb]
    \centering
    \includegraphics[width=1.0\linewidth]{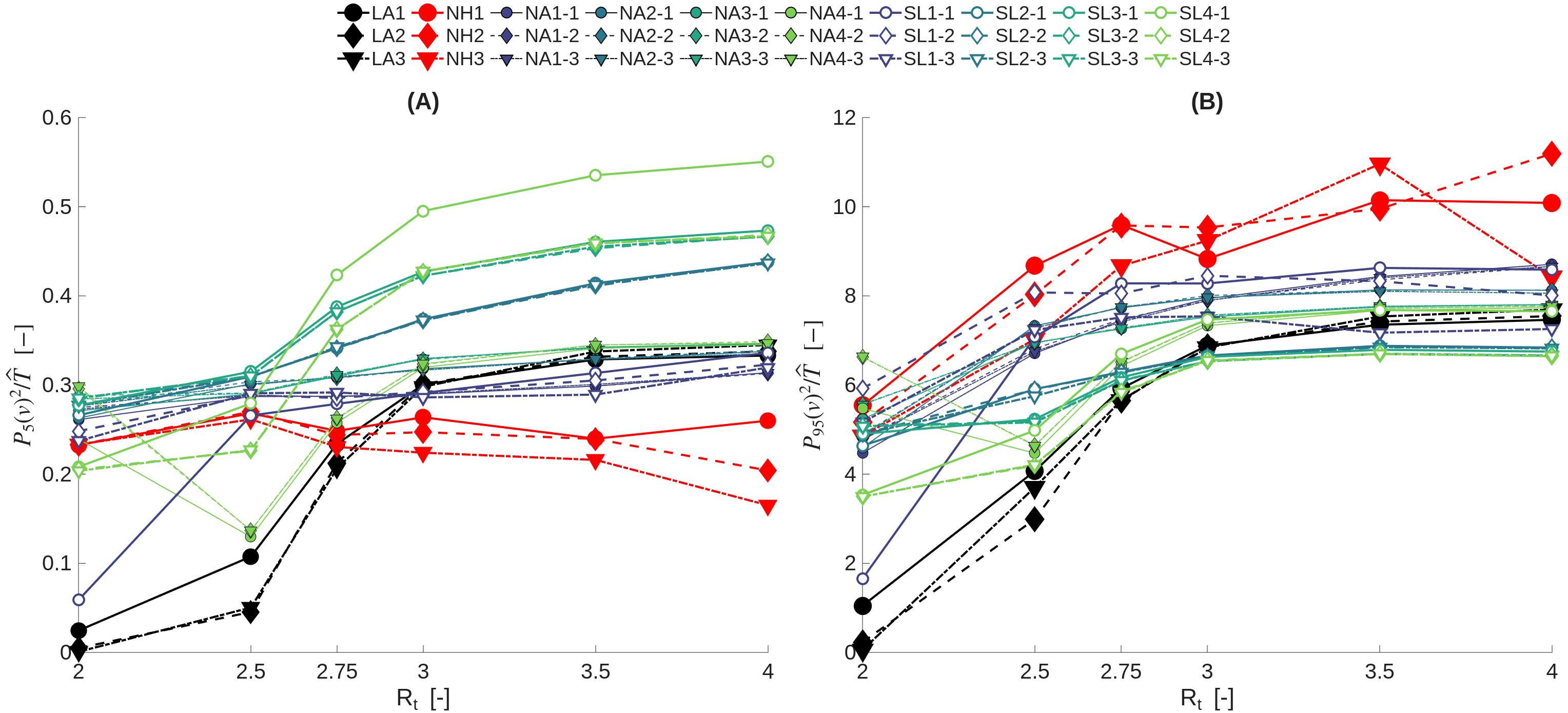}
    \caption{Effect of $R_t$ and thermostat coupling parameters ($\xi_\text{La}$ and $\xi_\text{NH}$) on (A) 5th and (B) 95th velocity percentiles, squared and normalised by prescribed temperature. }
    \label{fg_vel_prctiles_Rt}
\end{figure*}

\begin{figure}[!htb]
    \centering
    \includegraphics[width=\linewidth]{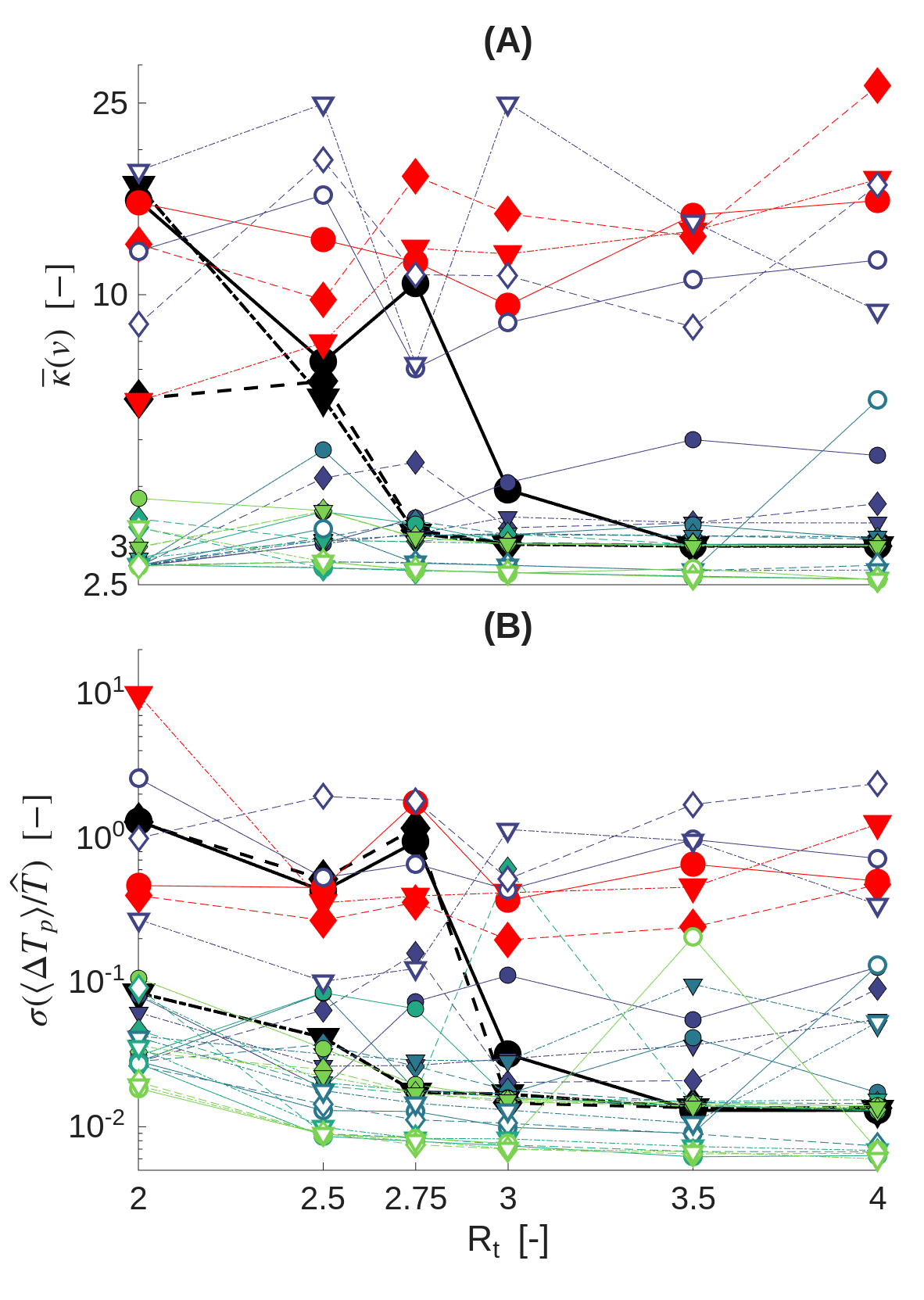}
    \caption{Effect of $R_t$, $\xi_\text{La}$ and $\xi_\text{NH}$ on (A) $\bar{\kappa}(v)$, (B) $\langle \Delta {T_p}\rangle$. The same legend of Figure \ref{fg_vel_prctiles_Rt} is employed.}
    \label{fg_granrand_std_Tp_Rt}
\end{figure}

\begin{figure*}[!htb]
    \centering
    \includegraphics[width=1.0\linewidth]{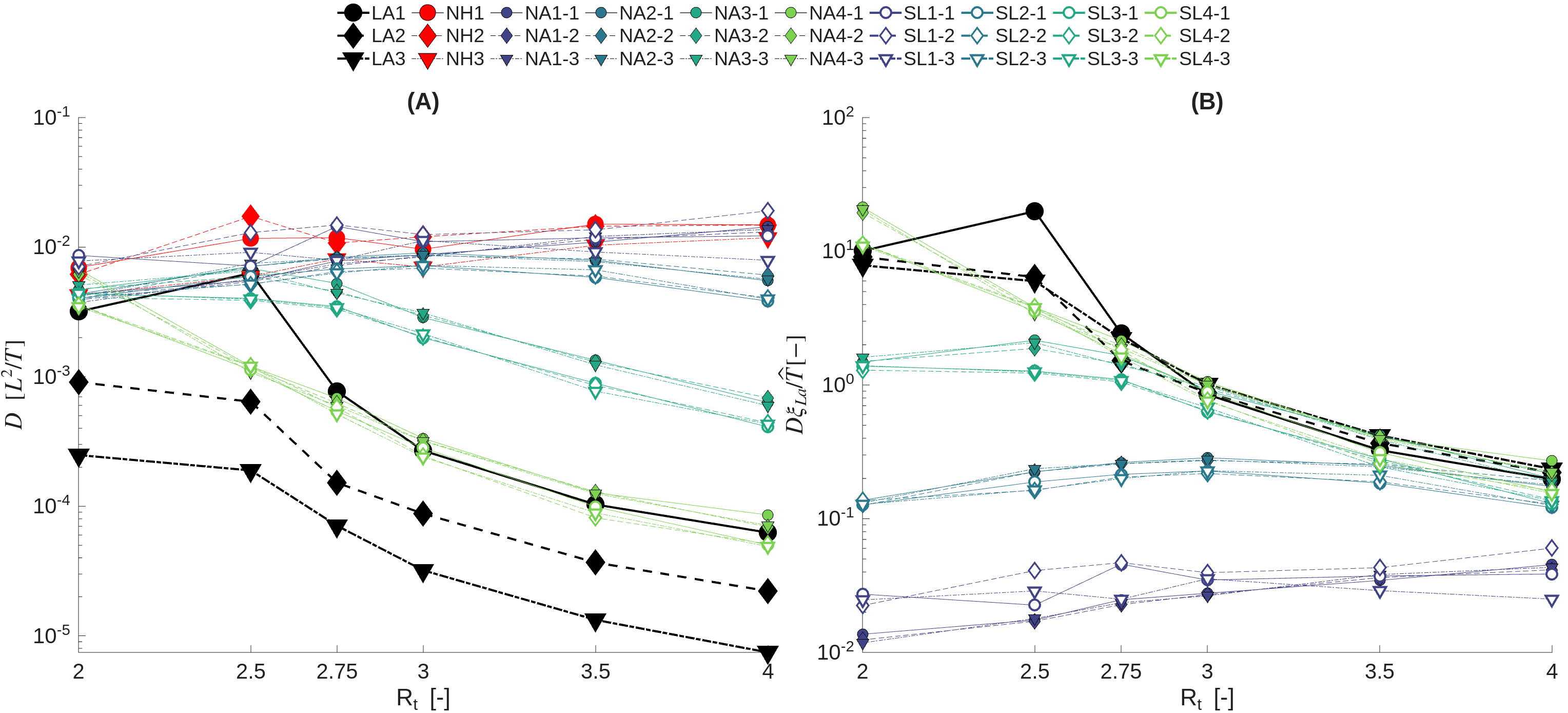}
    \caption{Effect of the thermostat parameters ($R_t$ and $\xi$) on diffusivity: (A) non-normalised, (B) normalised by $\xi_\text{La}/\hat{T}$.}
    \label{fg_granrand_D_Rt}
\end{figure*}

The following naming convention is used to summarise the results: the first two letters indicate the thermostat type (i.e. LA for Langevin, NH for Nos\'e-Hoover, NA for Naive Addition and SL for Scaled Langevin), while the number indicates progressively stronger coupling strength (see section \ref{sc_simulation_programme}). For hybrid formulations, the two numbers are separated by a dash. For example, NA1-3 indicates the naive implementation with $\xi_\text{La}=0.1$ and $\xi_\text{NH}=0.01$, while SL4-2 indicates the scaled Langevin approach with $\xi_\text{La}=100$ and $\xi_\text{NH}=0.1$. Since $\xi_\text{NH}$ often has no influence on the results, the second number is often omitted to indicate all simulations at a specific $\xi_\text{La}$ (e.g. NA1 corresponds to the group of NA1-1, NA1-2 and NA1-3). Figure \ref{fg_granrand_T_Rt} shows the time-averaged temperature for all cases. Subplot (A) corresponds to the individual particle velocity measurement ($\langle T\rangle$), while subplot (B) to the pair measurement ($\langle \Delta T_p\rangle$). As discussed in section \ref{sc_data_processing}, both metrics exhibit biases: $\langle T\rangle$ consistently overestimates the temperature in inhomogeneous scenarios where the assumption of zero center-of-mass velocity ($\boldsymbol{v}_g=0$) is violated (i.e. all NH cases and the SL cases with minimal stochastic kicks; SL1, $\xi_\text{La}=0.1$). 
Conversely, the pairwise approach $\langle T_p\rangle$ only measures the temperature in the portion of the domain affected by the thermostat, and therefore quantifies the local thermostat action. \\
\indent All thermostats with a deterministic component match the target ($\langle \Delta \bar{T}_p\rangle\approx0$), regardless of $\xi_\text{NH}$ and $R_t$. At equilibrium ($\Delta\zeta\approx0$) Nos\'e-Hoover injects the energy lost by inelastic collisions back into the system, enforcing the target enegy state ($\langle \bar T_p\rangle\approx\hat{T}$, $\bar\zeta\approx-\Gamma$, eq. (\ref{eq_Pv_NH})). However, the actual data is very noisy, and the emerging value depends greatly on the adopted sampling window and decisions (e.g. removing outliers significantly reduces the average value). Within the parameter space investigated, the temperature enforced by pure Langevin starts at $\langle \Delta \bar{T_p}\rangle/\hat{T}\approx-0.8$ ($R_t=2$) and only reaches $\langle \Delta \bar{T_p}\rangle/\hat{T}\approx-0.05$ for LA3 ($\xi_\text{La}=1000, R_t=4$).
\\ \indent The thermostat's efficacy at the pairwise scale, $\langle\bar{T}_p\rangle$, is partially reflected at the system scale, $\langle \bar{T}\rangle$. Given the reference packing density ($\phi\approx 0.27$) and interaction range ($R_t=2$), full pairwise thermalisation ($\langle\Delta\bar{T_p}\rangle\approx0$, as observed for all thermostats with a deterministic component), only corresponds to $\langle\bar{T}\rangle/\hat{T}\approx0.75$. The conversion rate between local (pairwise) and global (particle-wise) temperature improves until $R_t=3\sim3.5$. This emerging threshold matches the suggestion from \cite{ruiz2018effect}, to set $R_t$ as the first minimum of the radial distribution function ($ 1.5\sim1.75d = 3\sim3.5r_e$, see Figure \ref{fg_RDF_example}). The effect of the stochastic kicks are more complicated. Below the $R_t=3$ threshold, the stochastic term not only is unable to enforce the target temperature at the pair level ($\langle\Delta\bar T_p\rangle\ll0$ under pure Langevin), but it also hinders the transfer of kinetic energy from the thermostat-affected regions to the rest of the system \citep{garzo2004diffusion}. This effect can be observed in the hybrid thermostats plots: the plots LA1 (pure Langevin), NA4 and SL4 (hybrid), characterised by $\xi_\text{La}=100$, all exhibit roughly the same $\langle\bar{T}\rangle$ values within the $2.5\le R_t<3$ interval, despite the fact that the deterministic term enforces $\langle\Delta\bar{T}_p\rangle\approx0$ in the hybrid cases. We assume that this effect is not visible at $R_t=2$ because the low interaction range (and therefore low interaction time) limits the influence of the Langevin term. \\
\indent While the damping term ($-v_n\xi_\text{La}$) partially contributes to this effect (i.e. SL4 slightly over-performs NA4 in terms of $\langle\bar{T}\rangle/\hat{T}$ approaching 1), the stochastic kicks term appears to be the main cause: despite the fact that the SL simulations do not include the damping term, increasing $\xi_\text{La}$ still reduces $\langle\bar{T}\rangle$, (SL4$<$SL3$<$SL2). This behaviour is consistent with the previously observed trend of strong stochastic forcing reducing velocity correlations and therefore increasing kinetic energy dissipation even in the absence of an explicit damping term \citep{visco2007power,garzo2002transport}. \\
\indent Regarding pure Langevin simulations, an increase in $\xi_\text{La}$ produces a drop in $\langle\bar{T}\rangle$ below the interaction threshold, i.e. LA1$>$LA2$\simeq$LA3 for $R_t<3$. Above the $R_t$ threshold, the thermostat already acts homogeneously on the particles of the system and the role of heat diffusion disappears: the global temperature matches the pairwise value, $\langle\bar{T}_p\rangle\approx\langle\bar{T}\rangle$. \\
\indent Figure \ref{fg_vel_prctiles_Rt} shows the 5th (A) and 95th (B) percentiles of particle velocities. These metrics quantify: (i) the kinetic state of collisional cooling clusters ($P_5$); (ii) the population of high-velocity particles ($P_{95}$); and (iii) the degree of velocity segregation ($P_{95}-P_5$). Under pure Langevin, the two percentile plots replicate the $R_t$-dependent behaviour observed for the mean temperature values (Figure \ref{fg_granrand_T_Rt}A). For Nos\'e-Hoover, the reduced thermostat action at $R_t<3$ limits $P_{95}$ but does not affect $P_5$. The two hybrid thermostats exhibit significant differences: SL more strongly suppresses velocity segregation and collisional cooling, as indicated by its larger $P_5$ values, while NA produces statistics closer to pure Langevin.
\\ \indent Figure \ref{fg_granrand_std_Tp_Rt}, which assesses the thermostat precision, exhibits the same trends. 
Subplot (A) shows the time-averaged kurtosis of the particle velocities, $\bar{\kappa}(v)$, with $\kappa=3$ corresponding to a normal distribution. Subplot (B) shows the standard deviation of the domain-averaged pairwise temperature, $\sigma(\langle\Delta T_p\rangle/\hat{T})$, i.e. the amplitude of the temperature oscillations around the target. In all scenarios, pure Nos\'e-Hoover produces large oscillations, numerical instability, a highly leptokurtic ($\kappa\gg3$) distribution, and noisy results. Although both LA2 and LA3 exhibit the same degree of thermalisation (or lack thereof, e.g. $\langle\bar{T}\rangle\approx0$ at $R_t=2$), the stronger Langevin coupling of the latter significantly improves the thermostat precision (LA1 and LA2 exhibit $\sigma$ values one order of magnitude higher than LA3 under the same conditions). \\
\indent When the stochastic term is significant ($R_t\ge3,\xi_\text{La}\ge100$) both LA and NA exhibit the same velocity distribution ($\kappa$) and noise ($\sigma$). However, as soon as  $\xi_\text{La}$ drops below 100, the emerging velocity distribution has increased kurtosis (e.g. $\kappa>3$ in NA2 and NA3). This behaviour is not observed for SL, where the emerging distributions settle to $\kappa\approx2.75$, regardless of the stochastic kick magnitude (above 0.1). Additionally, due to linear scaling of the stochastic kick, which inhibits the development of high-velocity tails, SL is also characterised by the smallest amount of noise, with $\sigma$ reaching a minimum of $\approx 0.005$ for $\xi_\text{La}\ge10$. Thus, even when the scalar temperature is similarly controlled, the two hybrid formulations generate measurably different velocity statistics, confirming that temperature control and statistical decorrelation represent distinct aspects of the thermostatted state.
Figure \ref{fg_granrand_D_Rt} (A) shows the evolution of diffusivity, which matches the trends observed in the literature \citep{phan2013understanding,basconi_effects_2013,ruiz2018effect,braun2018anomalous}. The interaction range and the stochastic term have the most significant influence on diffusivity: for $\xi_\text{La}\ge10$ and $R_T\ge3$, $D\propto\xi_\text{La}^{-1}$. As shown in Figure \ref{fg_granrand_D_Rt} (B), normalising diffusivity as $D\xi_\text{La}/\hat{T}$ collapses LA, NA and SL onto a single curve. In all scenarios except $\xi_{La}=10^{-1}$ NA exhibits higher diffusivity than SL. This systematic difference is consistent with caging effects arising from dense packing ($\phi=0.27$, see Figure \ref{fg_RDF_example}): particles must overcome an energy barrier to escape their local neighbours and achieve a displacement greater than the mean free path. Therefore, the difference in diffusivity between NA and SL appears to be associated primarily with the upper portion of the velocity distribution ($P_{95}$ in Figure \ref{fg_vel_prctiles_Rt}B), rather than with their similar mean kinetic energy or with the kinetic state of the coldest particles. The Naive Addition approach applies a stochastic kick to particle pairs regardless of their current kinetic energy, thereby producing a higher $P_{95}$ and it is statistically more likely to overcome the caging barrier.\\
\indent While it is known that the interaction range affects diffusivity under deterministic temperature control \citep{ruiz2018effect,verbeek_advantages_2022}, the coupling parameter itself should have no influence at steady state. However, this is not necessarily the case here. Because granular materials are inherently out of equilibrium, the dynamics remain dependent on the thermostat stiffness outside the steady state. This was also observed in Figure \ref{fg_MSD_lag}. Highly segregated and non-ergodic systems exhibit temperature spikes due to inhomogeneous thermostat action and the emergence of power-law velocity tails. Under such conditions, an overly-stiff Nos\'e-Hoover formulation (NH3, $\xi_\text{NH}=0.01$) propagates the temperature oscillation in time \citep{hoover2004time}, reducing the overall diffusivity. This trend is visible in Figure \ref{fg_granrand_D_Rt} (A). For the hybrid formulation, the trend disappears (i.e. all curves with the same $\xi_\text{La}$ overlap), because even small stochastic contributions homogenise the velocity distribution enough to prevent the onset of large temperature fluctuations.
The final variable of interest is the particle distribution at contact $g(d)$, corresponding to the peak in the radial distribution function (Figure \ref{fg_RDF_example}). The reference value for elastic spheres is marked with a magenta line, with larger values indicating an increased probability of particles being in contact (i.e. clustering). Because Nos\'e-Hoover does not actively disrupt aggregation, it produces the highest degree of clustering ($\approx2.9$). Strong stochastic kicks ($\xi_\text{La}\ge100$) reduce clustering to the elastic baseline. Between these two extremes, the strength of the Langevin term controls the emerging $g(d)$. Surprisingly, pure Langevin simulations produce larger $g(d)$ values for $R_t>3$ and $\xi_\text{LA}>100$. We hypothesise that this is due to the stronger homogenisation effect breaking apart arching structures and reducing the overall local porosity, allowing particles to settle into denser packing \citep{knight1995density}.
\begin{figure}[!htb]
    \centering
    \includegraphics[width=\linewidth]{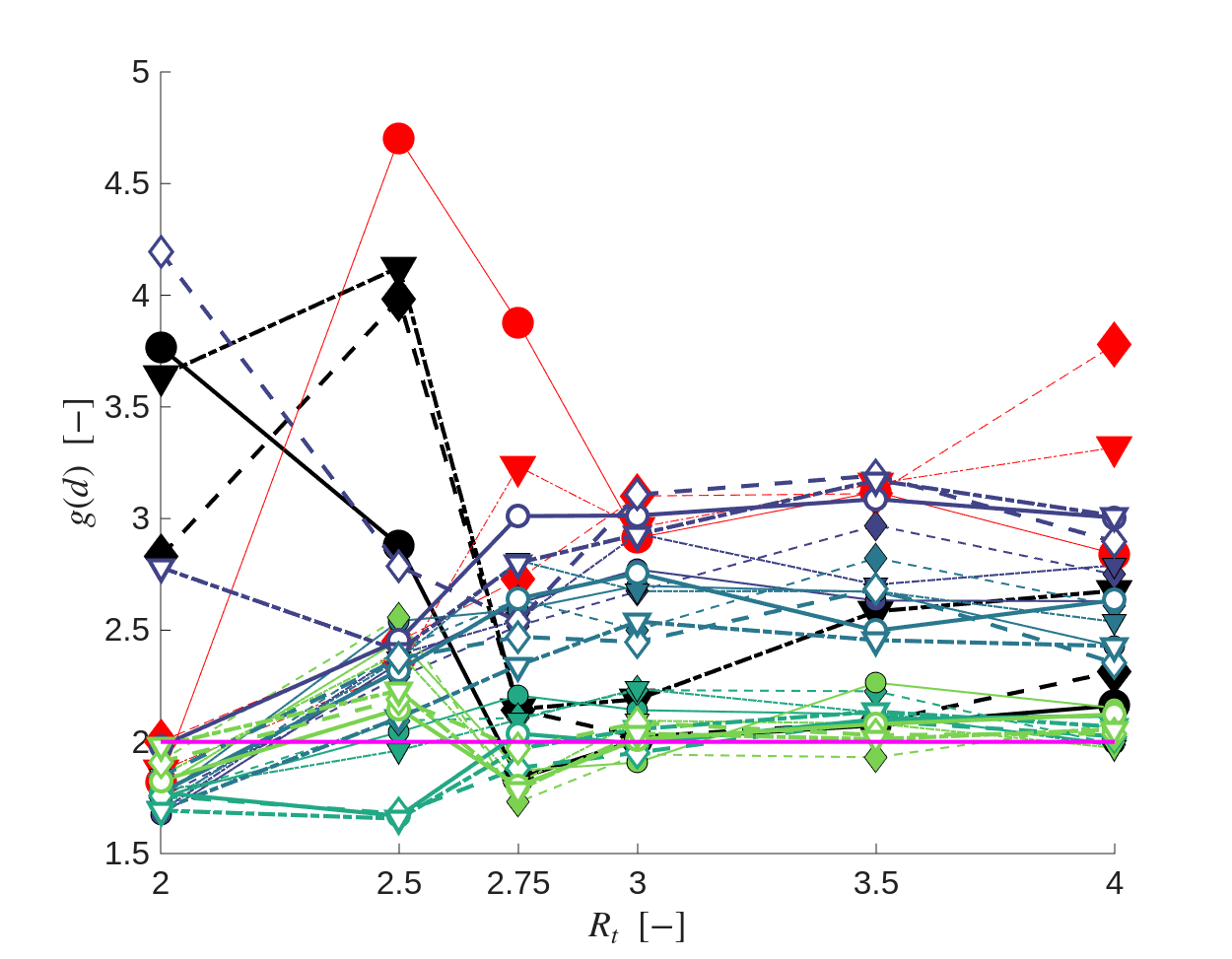}
    \caption{Effect of $R_t, \xi_\text{La}$ and $\xi_\text{NH}$ on $g(d)$. The same legend of Figure \ref{fg_granrand_D_Rt} is employed, except for the added magenta line at $g(d)=2$, which corresponds to the baseline value observed for fully elastic spheres. }
    \label{fg_g_d_Carnahan_Starling}
\end{figure}
\subsubsection{Study \#2, changing $\Delta t$, $\hat{T}$ and $\pi_2$}
\begin{figure*}[!htb]
    \centering
    \includegraphics[width=\linewidth]{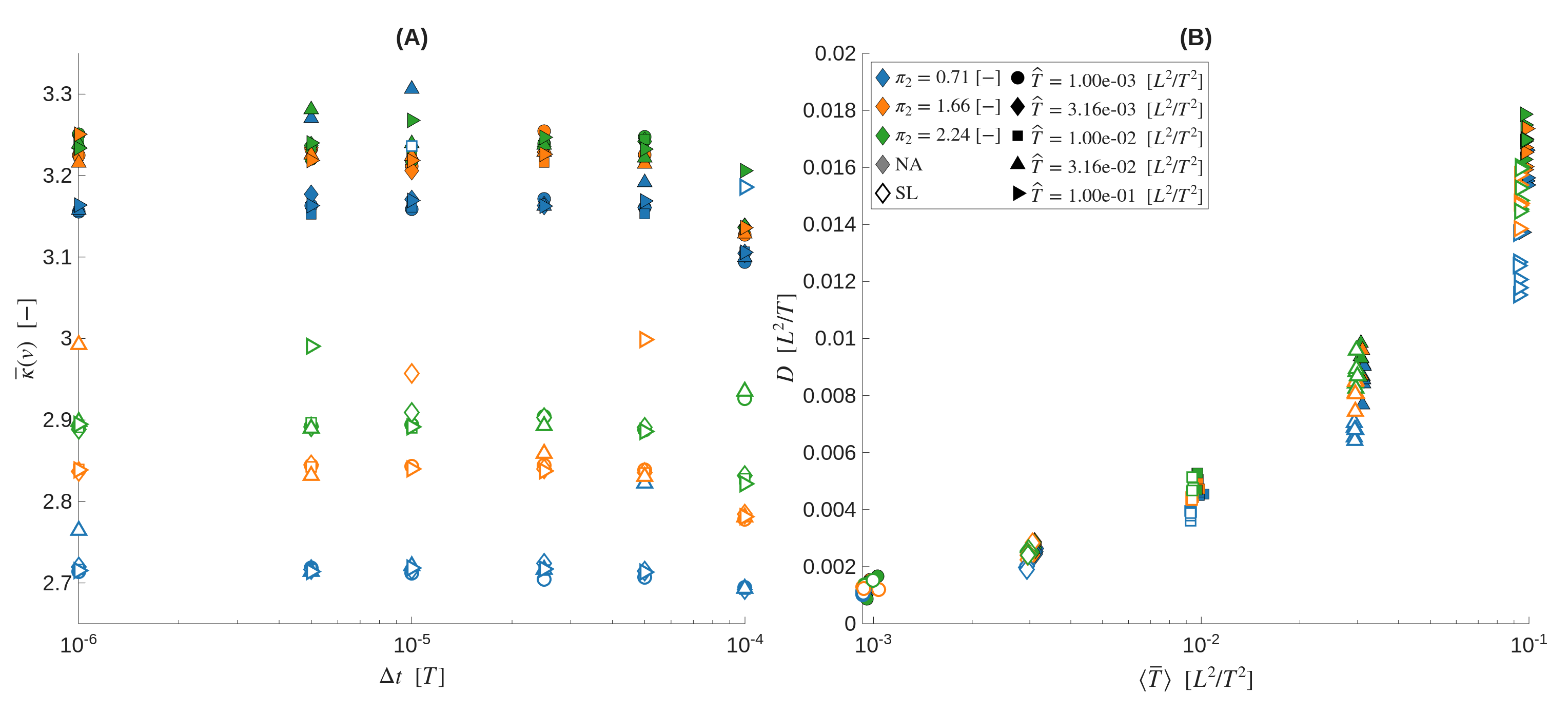}
    \caption{Numerical results from study \#2. (A) velocity kurtosis $\bar\kappa(v)$ is controlled by $\pi_2$ and thermostat formulation, and it is independent from $\Delta t$. (B) nonlinear relationship between $\langle\bar{T}\rangle$ and $D$. Diffusivity is firstly controlled by $\langle\bar{T}\rangle$, and secondarily by the thermostat formulation and $\pi_2$.}
    \label{fg_dt_kappa_T_D}
\end{figure*}

This study tests whether the thermostat remains applicable across a broad range of target temperatures and timestep sizes. Within the parameter space investigated, both SL and NA enforce the target temperature without inducing numerical instabilities. The results highlight three main findings. First, below $10^{-4}$, $\Delta t$ has a negligible influence on the meso-scale properties of the system, as shown in Figure \ref{fg_dt_kappa_T_D}A. Second, temperature and diffusivity exhibit a nonlinear relationship, i.e. $D\propto\langle\hat{T}\rangle^{0.5}$ (Figure \ref{fg_dt_kappa_T_D}B), matching the predictions of kinetic theory for a granular gas at the dilute limit \citep{brilliantov2001granular,garzo2002transport}. Third, the non-dimensional number $\pi_2$ governs the system response. This is shown using the time-averaged velocity kurtosis $\bar{\kappa}(v)$, which best captures the dependence on $\pi_2$ and thermostat formulation. The difference between NA and SL, already observed in the previous parametric study, reflects the different statistical action of their stochastic terms. Strong stochastic coupling drives NA toward an approximately mesokurtic ($\kappa=3$) Gaussian distribution, whereas the state-dependent forcing of SL produces a characteristic platykurtic distribution ($\kappa\approx2.6$). For both thermostats, $\pi_2$ controls the relative strength of the stochastic forcing. Lower values of $\pi_2$ apply a stronger stochastic term, which drives the velocity distribution toward the thermostat's prescribed shape. Higher values allow the system shift to revert toward the leptokurtic, non-Maxwellian distribution characteristic of highly segregated, pure Nos\'e-Hoover state ($\kappa(v)\gg3$). The same trend is observed in diffusivity, which increases with $\pi_2$.
\subsubsection{Study \#3, changing $e$, $\hat{T}$ and $\Delta t$}\label{sc_timestep_size}
\begin{figure*}[!htb]
    \centering
   \includegraphics[width=\linewidth]{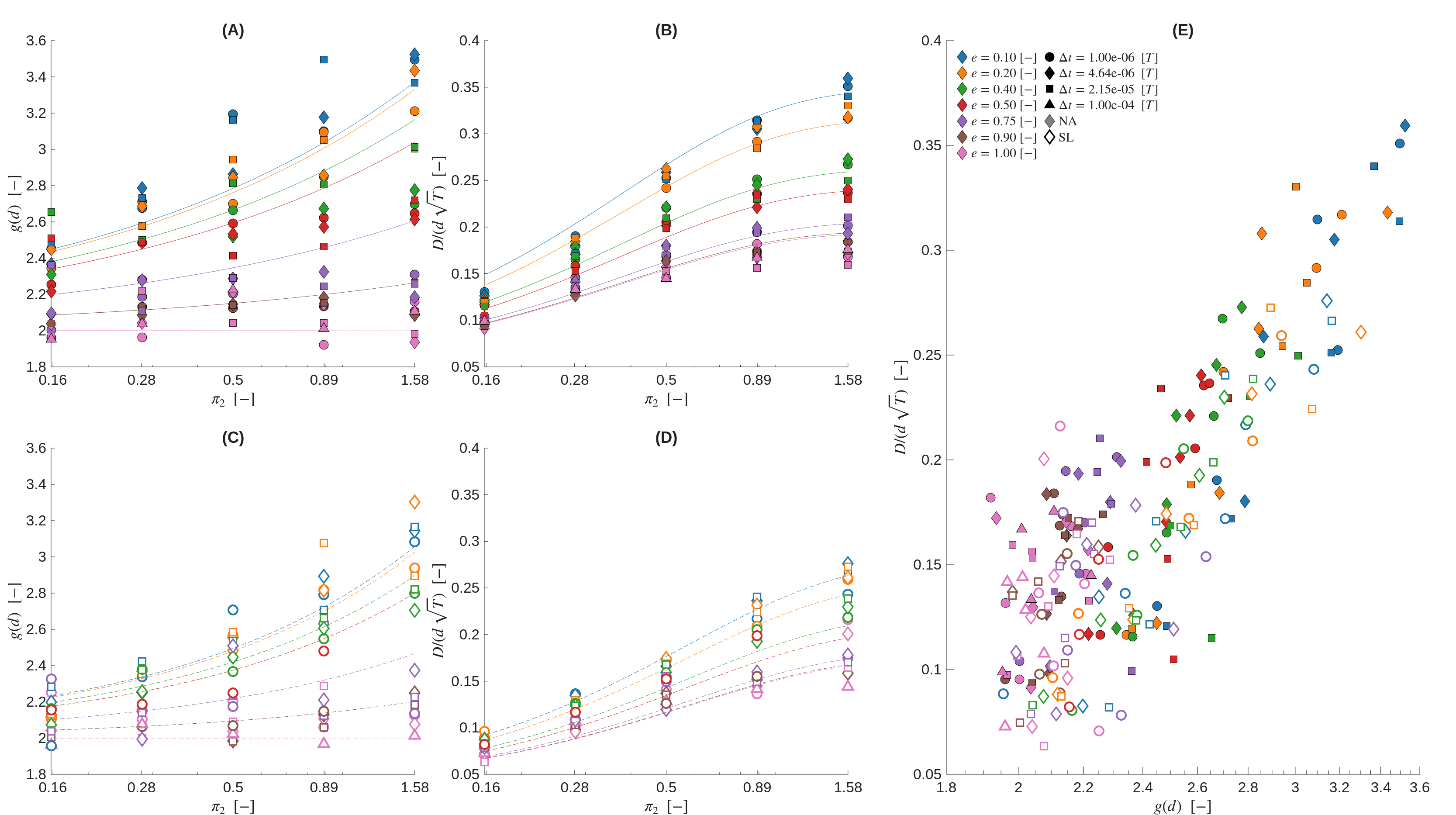}
    \caption{Emerging system properties from study \#3: (A) and (C) show $g(d)$ against $\pi_2$ for NA and SL, respectively; while (B) and (D) do the same for $D/(d\sqrt{\hat{T}})$. Subplot (E) shows the emerging relationship between $D/(d\sqrt{\hat{T}})$ and $g(d)$.}
    \label{fg_pi2_gd_D}
\end{figure*}

Figure \ref{fg_pi2_gd_D} shows the emerging system properties plotted against $\pi_2=\sqrt{\hat{T}}/(d\xi_\text{La})$. Fully elastic spheres ($e=1$) exhibit constant $g(d)\approx2$ regardless of the thermostat formulation and strength of the stochastic term (i.e. they are independent of $\pi_2$). As the degree of inelasticity increases, so does $g(d)$, and the system's dependence on $\pi_2$ becomes more significant. In fact, all emerging properties (e.g. $\kappa$, $P_{95}$, $g(d)$, $D$) can be captured using an expression of the form $y=y_{ref}+ax^b(1-e^2)$, where $y$ and $y_{ref}$ are respectively the given observables and their reference value for elastic particles and $x=\pi_2$ quantifies the relative timescale, and therefore the inverse strength, of the stochastic forcing. $a$ and $b$ are empirical fitting parameters that encapsulate the effect of parameters not investigated here (e.g. polydispersity, contact parameters, solid fraction). This form captures a linear dependence of the observed quantities on collisional energy-loss factor $(1-e^2)$, controlled by the nonlinear effect of the stochastic forcing strength $\pi_2$. This dependence is consistent with kinetic theory predictions for the effect of inelasticity on granular transport \citep{garzo2002transport,brilliantov2001granular}.\\
\indent Among these observables, $g(d)$ is particularly useful because, unlike velocity-distribution descriptors such as $\kappa$ and $P_{95}$, it appears to be broadly independent of the thermostat formulation (NA versus SL) and therefore provides a convenient mesoscopic descriptor of the state. This distinction indicates that different properties retain different degrees of sensitivity to the mechanism of stochastic energy injection.
Figure \ref{fg_pi2_gd_D}E shows a linear positive correlation between $D/(d{\hat{T}^{0.5}})$ and $g(d)$: states with stronger near-contact correlation also exhibit higher normalised diffusivity under the conditions of the study (fixed stochastic coupling and variable temperature and inelasticity). This behaviour matches previous observations by \cite{shah2025molecular} in a uniformly heated granular gas, where inelastic collisions led to higher diffusivity.
In light of these results, the hybrid formulations are relatively insensitive to the purely numerical control parameters within the investigated range. Neither the Nos\'e-Hoover coupling strength ($\xi_\text{NH}$) nor the timestep size ($\Delta t$) significantly affects the stationary state. Following \cite{ruiz2018effect}, the thermostat interaction range $R_t$ can be selected from the physical structure of the system and should approximately correspond to the first minimum of the RDF. The remaining stochastic coupling $\xi_\text{La}$ has a different role: through $\pi_2$, it controls the relative strength of stochastic decorrelation and therefore modifies the emerging velocity statistics, spatial structure and transport properties. The hybrid thermostat consequently separates the deterministic control of the mean fluctuation-energy scale ($\hat{T}$) from stochastic control of the statistical organisation of those fluctuations.

\section{Application example: simple shear}\label{sc_application}
\begin{figure}[!htb]
    \centering
    \includegraphics[width=\linewidth]{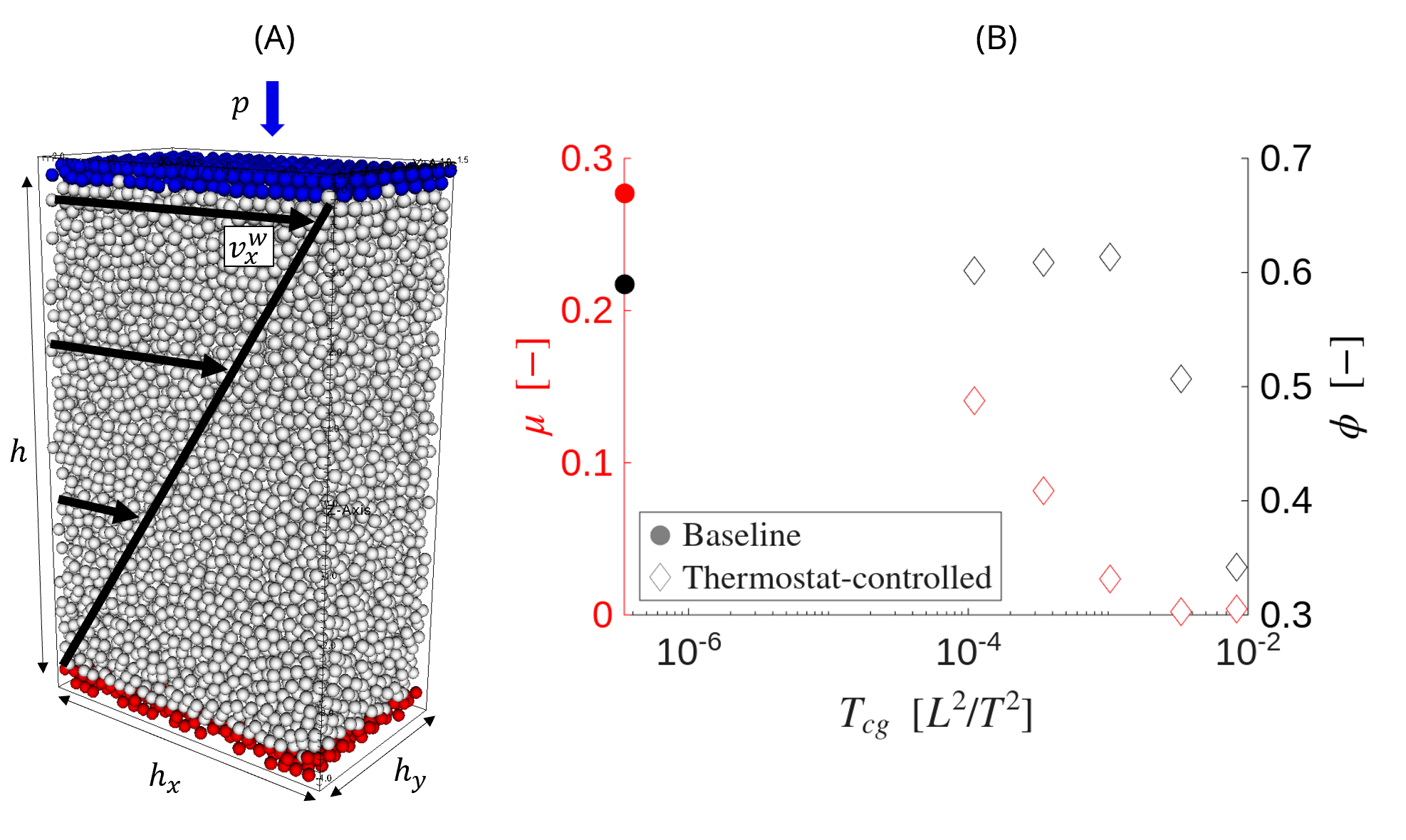}
    \caption{(A) Schematic representation of the shearbox, (B) Effect of granular temperature at constant inertia $I=1\cdot10^{-4}$ on the apparent friction and solid fraction of the ensemble.}
    \label{fg_T_mu_shearbox}
\end{figure}

In the following, we consider a standard boundary value problem used to investigate the rheology of granular materials \citep{gdr_midi_dense_2004} and, more recently, the effect of granular temperature therein \citep{irmer_granular_2024}, to illustrate the applicability of the developed pairwise hybrid thermostat. This boundary value problem is here implemented as a hexahedral domain with fixed periodic boundaries in the flow ($x$) and transversal ($y$) directions, and pressure-controlled bumpy walls in the shearing direction ($z$). The initial domain extents are respectively $h_x$ = 25$d_{max}$, $h_y$ = 15$d_{max}$ and $h$ = 30$d_{max}$. Particles are generated in a Hexagonal Close-Packed (HCP) configuration. After generation, 40\% of the particles are randomly deleted to prevent crystallisation and the sample is consolidated to the target pressure $p$ in the vertical ($z$) direction through servo-controlled horizontal frictionless walls. Once the model stabilises, all particles within a distance of $2d$ from the frictionless walls are fixed to obtain frictional, bumpy walls, which are then used to shear the sample. Throughout the simulation, the bottom boundary remains stationary, while the upper wall moves vertically to maintain the pressure constant, and horizontally to apply a constant shear strain rate. The strain rate is $\dot{\gamma}=v_x^w/h=0.001$, with $v_x^w$ being the horizontal velocity of the upper wall and $h$ the vertical extent at any point of the simulation, as shown in Figure \ref{fg_T_mu_shearbox}A. All DEM parameters are the same as in Table \ref{tb_dem_parameters}. Given constant vertical pressure ($p=12$, units of $M/(LT^2)$), the resulting inertial number is $I=1\cdot10^{-4}$ (dimensionless). Five different target temperatures above the unthermostatted baseline ($\approx 3.5\cdot10^{-7}$) are prescribed, i.e. $\hat{T}=[10^{-4},10^{-3.5},10^{-3},10^{-2.5},10^{-2}]$. The hybrid thermostat (NA) is used, with $\xi_\text{NH}=0.01$ and $\pi_2=1$. Since the upper boundary is pressure-controlled, the temperature affects the solid fraction $\phi$, which in turn influences the average particle distance spacing and therefore the interaction range $R_t$. Lacking an equation of state to predict $\phi$, we fix $R_t$ to a slightly larger value than suggested by the RDF plot in the previous study ($=3.5$). The temperature is sampled in the shearing ($z$) direction using coarse-grain statistics \citep{weinhart2012discrete} and then averaged over the central 80\% portion of the domain height. The two adopted output quantities are the apparent friction $\mu=\tau/p$ (where $\tau$ is the shear stress measured on the upper plate) and solid fraction $\phi$.\\
\indent Figure \ref{fg_T_mu_shearbox}B shows the resulting $\mu$ and $\phi$ as functions of the coarse-grained measured temperature $T_{cg}$. Overall, the thermostat maintains the target temperature reasonably well even under significant volumetric deformation, despite the associated changes in particle spacing and in the optimal thermostat interaction range $r_t$: for $\hat{T}=0.01$, $T_{cg}/\hat{T}\approx0.87$ and $\phi\approx0.34$. The measured friction broadly matches previously observed reduction in shear resistance with increasing granular temperature. At constant $I$, however, the resulting dependence does not follow the simple $\mu\propto\Theta^{-1/6}$ trend, indicating that additional state and material variables may remain relevant \citep{degiuli2016phase,vescovi2016merging,degiuli2017friction,man_friction-dependent_2022,irmer_granular_2024}. Similarly, the solid fraction $\phi$ presents a sharp transition between weak densification and significant dilation at $T_{cg}>10^{-3}$. The purpose of this example is not to establish a new constitutive scaling from a single fixed-$I$ series, but to demonstrate that granular temperature can be independently prescribed while quantities such as friction and solid fraction remain emergent properties of the controlled state.

\section{Summary and conclusions}
\indent This work investigates the applicability of thermostat algorithms to dissipative soft-sphere DEM. Conventional Langevin and Nos\'e-Hoover thermostats exhibit complementary limitations: Langevin requires strong coupling to compensate for collisional dissipation, thereby damping particle dynamics, whereas Nos\'e-Hoover enforces the mean target temperature without independently disrupting the velocity correlations and segregation generated by repeated inelastic collisions. Two hybrid formulations were therefore considered. Naive Addition (NA) combines conventional Langevin damping and stochastic forcing with Nos\'e-Hoover feedback, while Scaled Langevin (SL) removes the explicit Langevin damping and makes the stochastic forcing dependent on the instantaneous pair kinetic state. The two formulations consequently possess different statistical signatures: increasing stochastic forcing drives NA toward approximately mesokurtic velocity statistics, whereas SL produces a characteristic platykurtic distribution, despite both maintaining the same prescribed scalar temperature. In terms of energy-transfer mechanisms, NA is analogous to a homogeneous heat bath combined with feedback control, whereas SL is more closely related to mechanical forcing with a finite characteristic velocity. Thermostat properties are summarised in Table \ref{tb_thermostat properties}.
\begin{table*}[!htb]
\caption{\label{tb_thermostat properties}Summary of thermostat behaviour within the parameter range investigated. $^*$Controlled by $\pi_2$.}
\begin{ruledtabular}
\begin{tabular}{lcccc}
\textrm{Property}&
\textrm{Nos\'e-Hoover}&
\textrm{Langevin}&
\textrm{Naive Addition}&
\textrm{Scaled Langevin}\\
\colrule
Enforce $\langle \bar{T_p}\rangle\approx\hat{T}$  & YES & NO & YES & YES\\
Homogeneous in time & NO & YES & YES$^*$ & YES$^*$\\
Homogeneous in space & NO & YES & Partially$^*$ & Partially$^*$\\
Dissipate dynamics & NO & YES & YES$^*$ & YES$^*$\\
Velocity distribution & Leptokurtic & Mesokurtic & Mesokurtic$^*$ & Platykurtic$^*$\\
Physical analogue & Feedback controller & Heat bath & Heat bath  & Finite-velocity forcing\\
\end{tabular}
\end{ruledtabular}
\end{table*}
The parametric studies show that both hybrid formulations maintain the prescribed temperature with substantially weaker stochastic coupling than pure Langevin temperature control, while avoiding the strongly segregated and numerically unstable states observed under pure Nos\'e-Hoover. More importantly, prescribing temperature alone does not uniquely determine the kinetic state. The relative timescale of stochastic decorrelation is quantified by $\pi_2=\sqrt{\hat{T}}/(d\xi_\text{La})$, with lower $\pi_2$ corresponding to stronger stochastic forcing relative to particle motion. At fixed temperature, varying $\pi_2$ changes velocity statistics and diffusivity, demonstrating that transport depends independently on both fluctuation magnitude and stochastic decorrelation.\\
\indent The response to $\pi_2$ retains a dependence on the statistical character of the forcing. Higher-order velocity statistics and diffusivity differ between NA and SL, whereas the near-contact structure $g(d)$ approximately collapses onto a common $\pi_2$-$e$ relationship for the two formulations. Together with the observed dependence on the collisional energy-loss factor $(1-e^2)$, this suggests that stochastic decorrelation can be treated as an independent control variable for investigating granular transport and structure. Solid fraction, friction and polydispersity were not varied here, however, and must be incorporated before these empirical relationships can be interpreted as general constitutive closures.\\
\indent Finally, application to pressure-controlled simple shear demonstrates that granular temperature can be prescribed independently of the mechanically generated state, allowing exploration beyond the $T(I)$ trajectory produced naturally by shear and boundary work. The combined control of $\hat{T}$ and $\pi_2$, together with the distinct statistical signatures generated by different forcing operators, therefore provides a framework for identifying which kinetic and structural variables are required to close
temperature-dependent granular constitutive models. Experimentally or naturally forced states can be compared with these reference states by first matching $I$ and $T$ and then examining whether additional quantities such as velocity statistics, correlations and near-contact structure are also reproduced. Failure to obtain such equivalence would indicate that additional state variables are required.
\section*{CRediT authorship contribution statement}
Marco Previtali: Writing – review \& editing, Writing – original draft, Visualization, Validation, Methodology, Conceptualisation, Investigation, Software. Herbert Huppert: Writing - review \& editing. Sergio Andres Galindo-Torres: Writing – review \& editing, Supervision, Software, Conceptualisation, Methodology, Funding acquisition.
\section*{Declaration of competing interest}
The authors declare that they have no known competing financial interests or personal relationships that could have appeared to influence the work reported in this paper.
\section*{Acknowledgements}
We acknowledge the financial support from the National Natural Science Foundation of China with project number 12172305, Rheology and jamming transitions of dense granular-fluid systems. We thank Westlake University and the Westlake High-performance Computing Center for computational resources and corresponding assistance. 
\section*{Data availability}
All simulations were carried out using the open-source library Mechsys, available at https://mechsys.nongnu.org/. The raw data for the simulations presented in section \ref{sc_individual_comparison} is available at doi: 10.17632/yt9pvmvz9t.1. The full dataset (600 simulations, $\approx$ 200 GB) is available upon request.

\appendix
\section{Temperature control and velocity distributions under homogeneous constant cooling}
\subsection{Nos\'e-Hoover thermostat}\label{sc_nosehoover_Pv}
\renewcommand\theequation{\thesection\arabic{equation}}
\setcounter{equation}{0}
The emerging velocity distribution under pure Nos\'e-Hoover is obtained assuming spatial homogeneity and constant cooling $\Gamma$. The one-dimensional dynamics are given by eq. (\ref{eq_nose_hoover_partial_form_again})
 \begin{equation}\label{eq_nose_hoover_partial_form_again}
 \begin{aligned}[b]
     d v=-(\Gamma+\zeta)v\,dt\ \ \ \ \ \ &d\zeta=(v^2-\hat{T})/(Q\hat{T})\,dt.
   \end{aligned}
 \end{equation}
 Since both equations are deterministic, the corresponding phase-space drift velocities are $\dot v=-(\Gamma+\zeta)v$,$\dot\zeta=({v^2-\hat{T}})/({\hat{T}Q})$. When the system reaches the steady state, the velocity distribution is characterised by a joint probability density function $\rho(v,\zeta)$, which must satisfy the stationary Liouville equation, i.e. zero divergence in the $(v,\zeta)$ phase space \citep{tuckerman1999classical}
\begin{equation}\label{eq_NH_liouville_distribution}
     \frac{\partial}{\partial v}(\dot v\rho)+\frac{\partial}{\partial \zeta}(\dot\zeta\rho)=0.
 \end{equation}
 Substituting eq. (\ref{eq_nose_hoover_partial_form_again}) into eq. (\ref{eq_NH_liouville_distribution}), we obtain
 \begin{equation}\label{eq_NH_liouville_motion}
     \frac{\partial}{\partial v}[-(\Gamma+\zeta)v\rho]+\frac{\partial}{\partial \zeta}[(v^2-\hat{T})/(\hat{T}Q)\rho]=0.
 \end{equation}
A Gaussian trial joint distribution is proposed. Since eq. (\ref{eq_NH_liouville_motion}) is linear in $\zeta$ (except for $v^2$ in $\dot\zeta$), $\rho$ is proposed as an exponential of a quadratic (i.e. Gaussian) in $v$ and $\zeta$, with an added linear term in $\zeta$ to counteract the drift to zero induced by $\Gamma$
\begin{equation}\label{eq_ansatz_rho_NH}
    \rho(v,\zeta)=C\exp\left(-a/2v^2-b/2\zeta^2-c\zeta\right),
\end{equation}
where $a,b,c$ are unknown constants and $C$ is the normalisation factor, i.e. $\int_{-\infty}^{\infty}\int_{-\infty}^{\infty}\rho(v,\zeta)dvd\zeta=1$. The partial derivatives of eq. (\ref{eq_ansatz_rho_NH}) are given by
\begin{eqnarray}\label{eq_derivatives_ansatz_rho_NH}
         \partial \rho/\partial v=-av\rho,\ \ \ \ \ \ &\partial\rho/\partial\zeta=-(b\zeta+c)\rho.
\end{eqnarray}
The two terms of eq. (\ref{eq_NH_liouville_motion}) can then be computed as
\begin{equation}\label{eq_rho_NH_first}
\begin{split}
   \frac{\partial}{\partial v}[-(\Gamma+\zeta)v\rho]
   &=-(\Gamma+\zeta)\frac{\partial}{\partial v}(v\rho)\\
   &=-(\Gamma+\zeta)\left(\rho+v\frac{\partial\rho}{\partial v}\right)\\
   &=-(\Gamma+\zeta)(1-av^2)\rho.
\end{split}
\end{equation}
Likewise
\begin{equation}\label{eq_rho_NH_second}
\begin{split}
   \frac{\partial}{\partial \zeta}[(v^2-\hat{T})/(\hat{T}Q)\rho]
   &=[(v^2-\hat{T})/(\hat{T}Q)]\frac{\partial\rho}{\partial\zeta}\\
   &=(v^2-\hat{T})(-b\zeta-c)\rho/(\hat{T}Q).   
\end{split}
\end{equation}
Substituting into eq. (\ref{eq_NH_liouville_motion}) and dividing by $\rho$, we obtain
\begin{equation}\label{eq_NH_term_zero}
\begin{aligned}[b]
    -(\Gamma+\zeta)+a(\Gamma+\zeta)v^2 
    -(b\zeta+c)(v^2-\hat{T})/(\hat{T}Q)=0.\\
\end{aligned}
\end{equation}
The terms are then grouped by their coefficients ($v$ and $\zeta$) and each independent combination is set to zero. Since this equation must hold for arbitrary $v$ and $\zeta$, the independent coefficients give $a=1/\hat{T}$, $b=Q$, $c=\Gamma Q$.
The stationary joint distribution is therefore
\begin{equation}\label{eq_ansatz_rho_NH_solved}
\rho(v,\zeta)=C\exp[-{v^2}/({2\hat{T}})-{Q}\zeta^2/2
-\Gamma Q\zeta],
\end{equation}
Solving eq. (\ref{eq_ansatz_rho_NH_solved}), we obtain
\begin{equation}\label{eq_ansatz_rho_NH_solvedv2}
    \rho(v,\zeta)=C^\prime\exp[-v^2/(2\hat{T})-Q(\Gamma+\zeta)^2/2],
\end{equation}
where $C^\prime=C\exp[Q\Gamma^2/2]$. Therefore, the joint distribution $\rho$ is the product of two Gaussians, one in $v$ and the other in $\zeta$ (centered at $-\Gamma$). The latter can be marginalised by integrating over $\zeta$
\begin{eqnarray}
\label{eq_Pv_NH}
    P(v)=\int_{-\infty}^\infty\rho(v,\zeta)\,d\zeta\propto\exp[-v^2 
    /(2\hat{T})]\nonumber \\\int_{-\infty}^\infty\exp\left[-{Q}(\Gamma+\zeta)^2/2\right]\,d\zeta.
\end{eqnarray}
The second term is independent of $v$, so that $P(v)\propto\exp[-v^2/(2\hat{T})]$, with $\langle v^2\rangle=\hat{T}$. The stationary distribution of the Nos\'e-Hoover variable is therefore centred at $\langle \zeta\rangle=-\Gamma$.
\subsection{Naive Addition (Nos\'e-Hoover plus Langevin)}\label{sc_naive_addition_Pv}
We begin from the combined one-dimensional equation of motion
\begin{equation}\label{eq_combined}
\begin{aligned}
dv&=-(\Gamma+\xi_\text{La}+\zeta)v\,dt+\sqrt{2\xi_\text{La}\hat{T}}\,dW,\\
d\zeta&=(v^2-\hat{T})/(\hat{T}Q)\,dt,
\end{aligned}
\end{equation}
where $W(t)$ is the standard Wiener process, as defined in the main text. The stochastic equation is interpreted in the It\^{o} sense. For a stochastic differential equation of the form $dv=A_v\,dt+B\,dW$, the corresponding Fokker--Planck equation contains
a drift term $-\partial(A_v\rho)/\partial v$ and a diffusion term $\frac{1}{2}\partial^2(B^2\rho)/\partial v^2$. In the present case, $A_v=-(\Gamma+\xi_\text{La}+\zeta)v$ and the Langevin noise amplitude $B=\sqrt{2\xi_\text{La}\hat{T}}$ is independent of $v$, giving the diffusion coefficient $B^2/2=\xi_\text{La}\hat{T}$. Because the noise
is additive ($B$ is state independent), the It\^{o} and Stratonovich interpretations are equivalent for this formulation. The corresponding Fokker-Planck equation for the joint probability density $\rho(v,\zeta,t)$ is therefore \begin{equation}\label{eq_fokker_planck_combined}
\frac{\partial\rho}{\partial t}=-\frac{\partial}{\partial v}\left[-(\Gamma+\xi_\text{La}+\zeta)v\rho\right]-\frac{\partial}{\partial\zeta}\left[({v^2-\hat{T}})/({\hat{T}Q})\rho
\right]+\xi_\text{La}\hat{T}\frac{\partial^2\rho}{\partial v^2}.
\end{equation}
The distribution previously obtained under pure Nos\'e-Hoover is used as the trial solution for stationary conditions ($\partial\rho/\partial t=0$), to lead to
\begin{equation}\label{eq_ansatz_combined}
    \rho(v,\zeta)=C\exp[-v^2/(2\hat{T})-Q(\zeta+\Gamma)^2/2],
\end{equation}
with $C$ being the normalisation constant. The probability distribution is affected by probability currents $J_v$ (velocity changes) and $J_\zeta$ (changes in the Nos\'e-Hoover friction variable). $J_v$ is given by
\begin{equation}\label{eq_jv_current_combined}
    J_v=-(\Gamma+\xi_\text{La}+\zeta)v\rho-(\partial\rho/\partial v)(\xi_\text{La}\hat{T}).
\end{equation}
As before, the first term represents the velocity-space drift toward zero (due to collisional cooling, Langevin damping and Nos\'e-Hoover feedback) while the second term is the diffusive probability current associated to the stochastic contribution $\sqrt{2\xi_\text{La}\hat{T}}\,dW$. The $\xi_\text{La}$ terms cancel out when substituting $\partial\rho/\partial v=-(v/\hat{T})\rho$, resulting in $J_v=-(\Gamma+\zeta)v\rho$. As with pure Nos\'e-Hoover, the $J_\zeta$ current is the time-evolution of the $\zeta$ parameter due to the Nos\'e-Hoover thermostat, i.e. $d\zeta/dt=(v^2-\hat{T})/(\hat{T}Q)$ in eq. (\ref{eq_combined})
\begin{equation}\label{eq_jzeta_current_combined}
    J_\zeta=(v^2-\hat{T})/(\hat{T}Q)\rho.
\end{equation}
Since we aim for $\partial \rho/\partial t=0$, we calculate the divergences as
\begin{equation}\label{eq_div_jv_combined}
\begin{split}
\frac{\partial J_v}{\partial v}&=-\frac{\partial}{\partial v}[(\Gamma+\zeta)v\rho]\\
&=-(\Gamma+\zeta)\left[\rho+v\frac{\partial\rho}{\partial v}\right]\\
&=-(\Gamma+\zeta)(1-v^2/\hat{T})\rho,
\end{split}
\end{equation}
while
\begin{equation}\label{eq_div_jzeta_combined}
\begin{split}
\frac{\partial J_\zeta}{\partial \zeta}&=\frac{\partial\rho}{\partial \zeta}(v^2-\hat{T})/({\hat{T}Q})\\
&=-(\Gamma+\zeta)(v^2/\hat{T}-1)\rho.
\end{split}
\end{equation}
Summing and setting eq. (\ref{eq_div_jv_combined}) and (\ref{eq_div_jzeta_combined}) to zero, we obtain
\begin{equation}\label{eq_j_sum_current_combined}
\frac{\partial J_v}{\partial v}+\frac{\partial J_\zeta}{\partial \zeta}=-(\Gamma+\zeta)[(\hat{T}-v^2)/\hat{T}+(v^2-\hat{T})/\hat{T}]\rho=0.
\end{equation}
Given $(\hat{T}-v^2)+(v^2-\hat{T})=0$, the two $J$ terms cancel out, $\partial\rho/\partial t=0$ and eq. (\ref{eq_distribution_combined}) is the resulting velocity distribution, given by
 \begin{equation}\label{eq_distribution_combined}
\rho(v,\zeta)=\frac{1}{\sqrt{2\pi\hat{T}}}\sqrt{\frac{Q}{2\pi}}\exp\left[-\frac{v^2}{2\hat{T}}-\frac{Q}{2}(\zeta+\Gamma)^2\right].
\end{equation}
Integrating $\zeta$ to remove it returns the standard Maxwell-Boltzmann distribution
\begin{equation}\label{eq_distribution_combined_independent}
P(v)=\int_{-\infty}^{\infty}\rho(v,\zeta)\,d\zeta=
\exp[
-{v^2}/({2\hat{T})}
]/{\sqrt{2\pi\hat{T}}}.
\end{equation}
 Thus, under the homogeneous constant-cooling approximation, the Langevin damping and stochastic diffusion cancel out in the stationary probability current, while the stationary distribution of $\zeta$ remains centred at $\langle\zeta\rangle=-\Gamma$.
\subsection{Comparison with a vibrating boundary}
\label{sc_SL_vibrating_wall}
Consider a one-dimensional collision between a particle with pre-collisional velocity $v$ and a wall moving with instantaneous velocity $U$. The normal coefficient of restitution condition is $v'-U=-e_w(v-U)$, where $e_w$ is the particle-wall coefficient of restitution. The
post-collisional particle velocity is therefore $v'=(1+e_w)U-e_wv$ and the corresponding change in particle kinetic energy is
\begin{equation}\label{eq_wall_energy_change}
\begin{split}
    \Delta E
    &=
    {m}\left(v'^2-v^2\right)/2
    \\
    &=
    {m}
    \left[
        (1+e_w)^2U^2
        -2e_w(1+e_w)Uv
        -(1-e_w^2)v^2
    \right]/2.
\end{split}
\end{equation}
The actual energy-transfer statistics of a vibrating boundary depend on the wall phase at collision and on correlations between $U$ and the incoming particle velocity. A simple limiting case is obtained by assuming a symmetric wall-velocity distribution with $\langle U\rangle=0$ and neglecting correlations between $U$ and $v$. Equation (\ref{eq_wall_energy_change}) then gives $\langle\Delta E|v\rangle =m[(1+e_w)^2\langle U^2\rangle-(1-e_w^2)v^2 ]/2$. Thus, even under fixed vibration conditions, the mean energy supplied by the wall is not independent of the incoming particle velocity. It decreases with $v^2$ and changes sign at the characteristic velocity $v_c^2 =(1+e_w)/(1-e_w)   \langle U^2\rangle$.

For the sinusoidal motion typically considered in vibrating boundaries, e.g. \cite{irmer_granular_2024}, $U=V_0\cos(\omega t)$ with $V_0=A\omega$, uniform sampling of the vibration phase would give $\langle U^2\rangle=V_0^2/2$, and therefore
\begin{equation}\label{eq_wall_characteristic_velocity_sinusoidal}
    v_c = V_0\sqrt{(1+e_w)/[2(1-e_w)]}.
\end{equation}

\section{Assessment of viscosity from equilibrium fluctuation methods}\label{sc_viscosity}
\begin{figure}[!b]
    \centering
    \includegraphics[width=\linewidth]{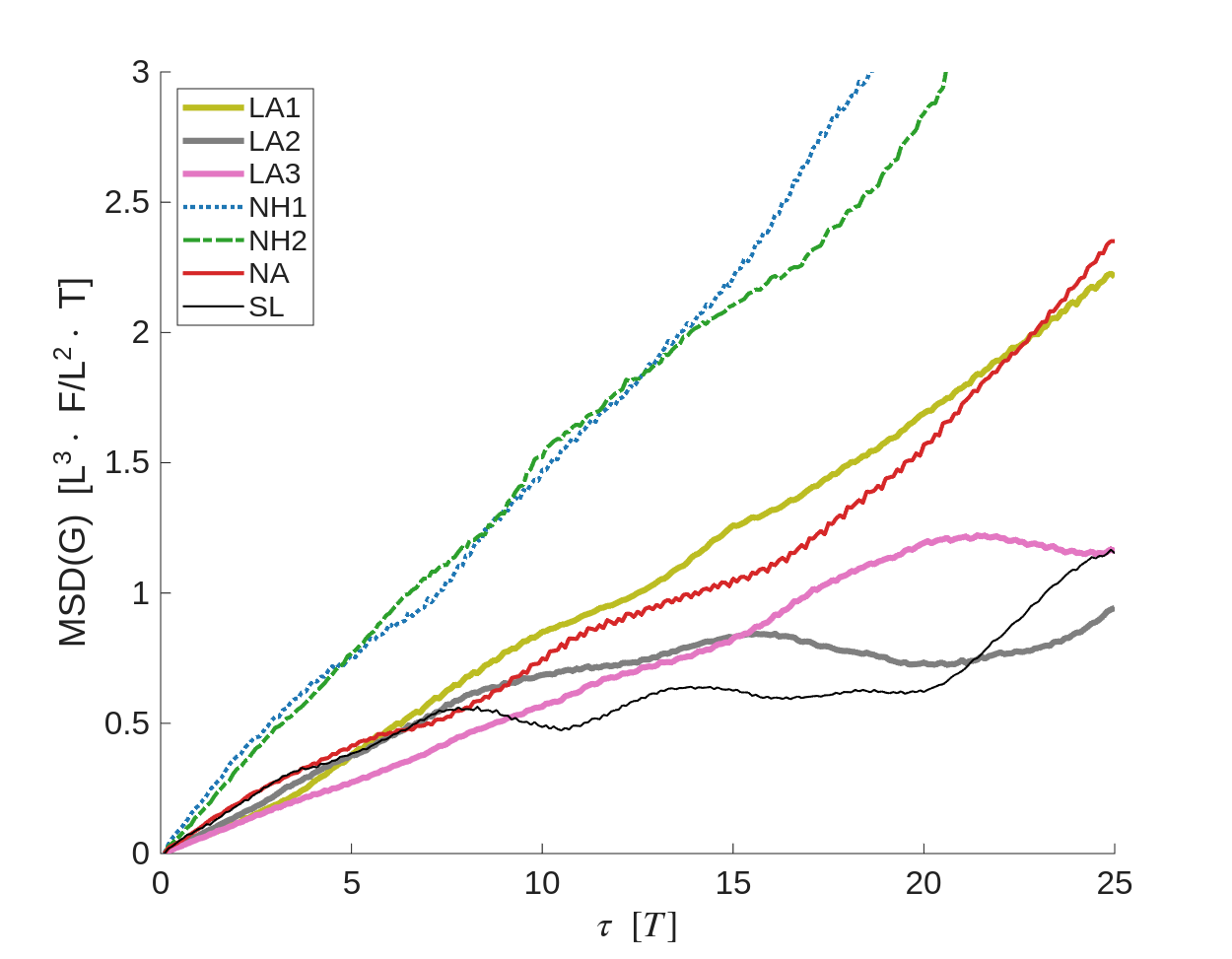}
    \caption{MSD(G) plots for the different thermostats. LA1, $\xi_\text{La}=10^2,\ R_t=3$; LA2 $\xi_\text{La}=10^3,\ R_t=3$; LA3, $\xi_\text{La}=10^3,\ R_t=4$. NH1 $\xi_\text{NH}=10^{-2}$, NH2, $\xi_\text{NH}=10^0$. NA and SL have the same parameters ($\xi_\text{La}=10^0,\ \xi_\text{NH}=10^{-2}$).}
    \label{fg_granrand_MSDG_Lag}
\end{figure}

For completeness, the tentative viscosity calculation used in this paper is reported below. As for diffusivity, viscosity can be evaluated via the Green-Kubo relation, or, as is typically done for robustness, through the Einstein-Helfand method \citep{helfand1960transport,viscardy2007transport}. The Helfand moment $G_{ij}(t)$ is defined as the cumulative sum of the instantaneous shear stress components
 \begin{equation}\label{eq_helfand_moment}
 G_{ij}(t)=V\int_{t_0}^t\sigma_{ij}(t')dt',
 \end{equation}
 where $V$ is the volume of the system. The shear viscosity $\eta_{ij}$ is obtained from the long-time slope of the mean-square displacement of $G_{ij}$
 \begin{equation}\label{eq_MSD_G}
     \eta_{ij} = \lim_{\tau\rightarrow\infty}\overline{[G_{ij}(t+\tau)-G_{ij}(t)]^2}/(2VT\tau).
 \end{equation}
Figure \ref{fg_granrand_MSDG_Lag} shows the evolution of MSD(G) for the different thermostats. The growth is only linear for a limited lag ($\tau<10$), preventing the extrapolation of a steady-state viscosity coefficient for $\tau\rightarrow\infty$. Within this short interval, deterministic temperature control (Nos\'e-Hoover) results in higher apparent viscosity than stochastic approaches. Unlike in \cite{stoyanov_molecular_2005}, where viscosity emerged as a function of the probability of application of the stochastic perturbation, here even a weak stochastic term ($\xi_\text{La}=1.0$) causes a significant drop in the short-time response due to the disruption of frictional chains. The strength of the coupling parameter itself does not appear to have a large influence under either pure Nos\'e-Hoover or pure Langevin, e.g. LA1 $\simeq$ LA2 and NH1 $\simeq$ NH2, while the interaction range seems to further reduce the apparent viscosity for strong Langevin coupling (e.g. LA3).
No systematic correlation was observed between viscosity and the thermostat parameters (Study \#1). The emerging $\eta$ values varied by more than an order of magnitude without any clear dependence on $R_t$ and $\xi_\text{La}$. The chaotic behaviour of the observed MSD(G) trends suggests that the system's mechanical response cannot be described by a linear viscosity framework. While this failure is consistent with the presence of non-Markovian effects \citep{campbell2002granular} as it is typically expected in granular systems \citep{goldhirsch2000green}, the chaotic MSD(G) behaviour was also observed under strong stochastic forcing (LA), suggesting the persistence of frictional contact chains under large stochastic forcing.

\bibliography{bibliography}

@article{fei2026scaling,
  title={Scaling laws for vibration-induced friction weakening of quasi-statically sheared granular assembly},
  author={Fei, Jianbo and Tang, Hao and Chen, Xiangsheng},
  journal={Acta Geotechnica},
  pages={1--22},
  year={2026},
  publisher={Springer}
}

@article{alessio2026dense,
  title={Dense granular rheology from fluctuations},
  author={Alessio, Benjamin M and Edwards, Matthew R and Lai, Ching-Yao},
  journal={arXiv preprint arXiv:2601.01907},
  year={2026}
}

@article{pouliquen2026non,
  title={Non-local rheology in granular media: a perspective on the 2015 EPJE Paper by Bouzid et al.},
  author={Pouliquen, Olivier},
  journal={The European Physical Journal E},
  volume={49},
  number={4},
  pages={32},
  year={2026},
  publisher={Springer}
}

@article{windows2013thermal,
  title={Thermal convection and temperature inhomogeneity in a vibrofluidized granular bed: the influence of sidewall dissipation},
  author={Windows-Yule, CRK and Rivas, N and Parker, DJ},
  journal={Physical review letters},
  volume={111},
  number={3},
  pages={038001},
  year={2013},
  publisher={APS}
}

@article{basconi_effects_2013,
	title = {Effects of {Temperature} {Control} {Algorithms} on {Transport} {Properties} and {Kinetics} in {Molecular} {Dynamics} {Simulations}},
	volume = {9},
	issn = {1549-9618, 1549-9626},
	url = {https://pubs.acs.org/doi/10.1021/ct400109a},
	doi = {10.1021/ct400109a},
	language = {en},
	number = {7},
	urldate = {2025-08-12},
	journal = {Journal of Chemical Theory and Computation},
	author = {Basconi, Joseph E. and Shirts, Michael R.},
	month = jul,
	year = {2013},
	pages = {2887--2899}
}

@article{brilliantov1996model,
  title={Model for collisions in granular gases},
  author={Brilliantov, Nikolai V and Spahn, Frank and Hertzsch, Jan-Martin and P{\"o}schel, Thorsten},
  journal={Physical review E},
  volume={53},
  number={5},
  pages={5382},
  year={1996},
  publisher={APS}
}

@article{galindo2013coupled,
  title={A coupled discrete element lattice Boltzmann method for the simulation of fluid--solid interaction with particles of general shapes},
  author={Galindo-Torres, SA},
  journal={Computer Methods in Applied Mechanics and Engineering},
  volume={265},
  pages={107--119},
  year={2013},
  publisher={Elsevier}
}

@article{cundall1979discrete,
  title={A discrete numerical model for granular assemblies},
  author={Cundall, Peter A and Strack, Otto DL},
  journal={geotechnique},
  volume={29},
  number={1},
  pages={47--65},
  year={1979},
  publisher={Thomas Telford Ltd}
}

@article{degiuli2016phase,
  title={Phase diagram for inertial granular flows},
  author={DeGiuli, E and McElwaine, JN and Wyart, M},
  journal={Physical Review E},
  volume={94},
  number={1},
  pages={012904},
  year={2016},
  publisher={APS}
}

@article{leimkuhler2009gentle,
  title={A gentle stochastic thermostat for molecular dynamics},
  author={Leimkuhler, Ben and Noorizadeh, Emad and Theil, Florian},
  journal={Journal of Statistical Physics},
  volume={135},
  number={2},
  pages={261--277},
  year={2009},
  publisher={Springer}
}

@article{koopman2006advantages,
  title={Advantages of a Lowe-Andersen thermostat in molecular dynamics simulations},
  author={Koopman, EA and Lowe, CP},
  journal={The Journal of chemical physics},
  volume={124},
  number={20},
  year={2006},
  publisher={AIP Publishing}
}

@article{kim_power-law_2020,
	title = {Power-{Law} {Scaling} in {Granular} {Rheology} across {Flow} {Geometries}},
	volume = {125},
	issn = {0031-9007, 1079-7114},
	url = {https://link.aps.org/doi/10.1103/PhysRevLett.125.088002},
	doi = {10.1103/PhysRevLett.125.088002},
	language = {en},
	number = {8},
	urldate = {2025-08-12},
	journal = {Physical Review Letters},
	author = {Kim, Seongmin and Kamrin, Ken},
	month = aug,
	year = {2020},
	pages = {088002}
}

@article{leimkuhler2016pairwise,
  title={Pairwise adaptive thermostats for improved accuracy and stability in dissipative particle dynamics},
  author={Leimkuhler, Benedict and Shang, Xiaocheng},
  journal={Journal of Computational Physics},
  volume={324},
  pages={174--193},
  year={2016},
  publisher={Elsevier}
}

@article{leimkuhler2015numerical,
  title={On the numerical treatment of dissipative particle dynamics and related systems},
  author={Leimkuhler, Benedict and Shang, Xiaocheng},
  journal={Journal of Computational Physics},
  volume={280},
  pages={72--95},
  year={2015},
  publisher={Elsevier}
}

@article{qian2009effective,
  title={Effective control of the transport coefficients of a coarse-grained liquid and polymer models using the dissipative particle dynamics and Lowe--Andersen equations of motion},
  author={Qian, Hu-Jun and Liew, Chee Chin and M{\"u}ller-Plathe, Florian},
  journal={Physical Chemistry Chemical Physics},
  volume={11},
  number={12},
  pages={1962--1969},
  year={2009},
  publisher={Royal Society of Chemistry}
}

@article{berzi_granular_2024,
	title = {On granular flows: {From} kinetic theory to inertial rheology and nonlocal constitutive models},
	volume = {9},
	issn = {2469-990X},
	shorttitle = {On granular flows},
	url = {https://link.aps.org/doi/10.1103/PhysRevFluids.9.034304},
	doi = {10.1103/PhysRevFluids.9.034304},
	language = {en},
	number = {3},
	urldate = {2025-08-12},
	journal = {Physical Review Fluids},
	author = {Berzi, Diego},
	month = mar,
	year = {2024},
	pages = {034304}
}

@article{goldhirsch2008introduction,
  title={Introduction to granular temperature},
  author={Goldhirsch, Isaac},
  journal={Powder technology},
  volume={182},
  number={2},
  pages={130--136},
  year={2008},
  publisher={Elsevier}
}

@article{garzo2007enskog,
  title={Enskog theory for polydisperse granular mixtures. I. Navier-Stokes order transport},
  author={Garz{\'o}, Vicente and Dufty, James W and Hrenya, Christine M},
  journal={Physical Review E—Statistical, Nonlinear, and Soft Matter Physics},
  volume={76},
  number={3},
  pages={031303},
  year={2007},
  publisher={APS}
}

@book{phan2013understanding,
  title={Understanding viscoelasticity: an introduction to rheology},
  author={Phan-Thien, Nhan and Mai-Duy, Nam},
  year={2013},
  publisher={Springer}
}

@article{li2019influence,
  title={Influence of thermostatting on nonequilibrium molecular dynamics simulations of heat conduction in solids},
  author={Li, Zhen and Xiong, Shiyun and Sievers, Charles and Hu, Yue and Fan, Zheyong and Wei, Ning and Bao, Hua and Chen, Shunda and Donadio, Davide and Ala-Nissila, Tapio},
  journal={The Journal of chemical physics},
  volume={151},
  number={23},
  year={2019},
  publisher={AIP Publishing}
}

@article{braun2018anomalous,
  title={Anomalous effects of velocity rescaling algorithms: the flying ice cube effect revisited},
  author={Braun, Efrem and Moosavi, Seyed Mohamad and Smit, Berend},
  journal={Journal of chemical theory and computation},
  volume={14},
  number={10},
  pages={5262--5272},
  year={2018},
  publisher={ACS Publications}
}

@article{martyna_nosehoover_1992,
	title = {Nosé–{Hoover} chains: {The} canonical ensemble via continuous dynamics},
	volume = {97},
	issn = {0021-9606, 1089-7690},
	shorttitle = {Nosé–{Hoover} chains},
	url = {https://pubs.aip.org/jcp/article/97/4/2635/927962/Nose-Hoover-chains-The-canonical-ensemble-via},
	doi = {10.1063/1.463940},
	language = {en},
	number = {4},
	urldate = {2025-08-12},
	journal = {The Journal of Chemical Physics},
	author = {Martyna, Glenn J. and Klein, Michael L. and Tuckerman, Mark},
	month = aug,
	year = {1992},
	pages = {2635--2643}
}

@article{weinhart2012discrete,
  title={From discrete particles to continuum fields near a boundary},
  author={Weinhart, Thomas and Thornton, Anthony R and Luding, Stefan and Bokhove, Onno},
  journal={Granular Matter},
  volume={14},
  number={2},
  pages={289--294},
  year={2012},
  publisher={Springer}
}

@article{prevost2002forcing,
  title={Forcing and velocity correlations in a vibrated granular monolayer},
  author={Prevost, Alexis and Egolf, David A and Urbach, Jeffrey S},
  journal={Physical review letters},
  volume={89},
  number={8},
  pages={084301},
  year={2002},
  publisher={APS}
}

@article{reis2007forcing,
  title={Forcing independent velocity distributions in an experimental granular fluid},
  author={Reis, Pedro M and Ingale, Rohit A and Shattuck, Mark D},
  journal={Physical Review E—Statistical, Nonlinear, and Soft Matter Physics},
  volume={75},
  number={5},
  pages={051311},
  year={2007},
  publisher={APS}
}

@article{adachi2019magnetic,
  title={Magnetic excitation of a granular gas as a bulk thermostat},
  author={Adachi, Masato and Yu, Peidong and Sperl, Matthias},
  journal={npj Microgravity},
  volume={5},
  number={1},
  pages={19},
  year={2019},
  publisher={Nature Publishing Group UK London}
}

@article{gonzalez2022kinetic,
  title={Kinetic theory of granular particles immersed in a molecular gas},
  author={Gonz{\'a}lez, Rub{\'e}n G{\'o}mez and Garz{\'o}, Vicente},
  journal={Journal of Fluid Mechanics},
  volume={943},
  pages={A9},
  year={2022},
  publisher={Cambridge University Press}
}

@article{gomez2023diffusion,
  title={Diffusion of intruders in granular suspensions: Enskog theory and random walk interpretation},
  author={G{\'o}mez Gonz{\'a}lez, Rub{\'e}n and Abad, Enrique and Bravo Yuste, Santos and Garz{\'o}, Vicente},
  journal={Physical Review E},
  volume={108},
  number={2},
  pages={024903},
  year={2023},
  publisher={APS}
}

@article{kamrin_nonlocal_2012,
	title = {Nonlocal {Constitutive} {Relation} for {Steady} {Granular} {Flow}},
	volume = {108},
	copyright = {http://link.aps.org/licenses/aps-default-license},
	issn = {0031-9007, 1079-7114},
	url = {https://link.aps.org/doi/10.1103/PhysRevLett.108.178301},
	doi = {10.1103/PhysRevLett.108.178301},
	language = {en},
	number = {17},
	urldate = {2025-08-12},
	journal = {Physical Review Letters},
	author = {Kamrin, Ken and Koval, Georg},
	month = apr,
	year = {2012}
}

@article{patra_nonergodicity_2014,
	title = {Nonergodicity of the {Nose}-{Hoover} chain thermostat in computationally achievable time},
	volume = {90},
	copyright = {http://link.aps.org/licenses/aps-default-license},
	issn = {1539-3755, 1550-2376},
	url = {https://link.aps.org/doi/10.1103/PhysRevE.90.043304},
	doi = {10.1103/PhysRevE.90.043304},
	language = {en},
	number = {4},
	urldate = {2025-08-12},
	journal = {Physical Review E},
	author = {Patra, Puneet Kumar and Bhattacharya, Baidurya},
	month = oct,
	year = {2014},
	pages = {043304}
}

@article{mcnamara1997energy,
  title={Energy flux into a fluidized granular medium at a vibrating wall},
  author={McNamara, Sean and Barrat, Jean-Louis},
  journal={Physical Review E},
  volume={55},
  number={6},
  pages={7767},
  year={1997},
  publisher={APS}
}

@misc{man_friction-dependent_2022,
	title = {Friction-dependent rheology of dry granular systems},
	url = {http://arxiv.org/abs/2205.14898},
	doi = {10.48550/arXiv.2205.14898},
	language = {en},
	urldate = {2025-08-12},
	publisher = {arXiv},
	author = {Man, Teng and Zhang, Pei and Ge, Zhuan and Galindo-Torres, Sergio A. and Hill, Kimberly M.},
	month = may,
	year = {2022},
	note = {arXiv:2205.14898 [cond-mat]}
}

@misc{allen_thermostat_2006,
	title = {A {Thermostat} for {Molecular} {Dynamics} of {Complex} {Fluids}},
	url = {http://arxiv.org/abs/cond-mat/0606511},
	doi = {10.48550/arXiv.cond-mat/0606511},
	language = {en},
	urldate = {2025-08-12},
	publisher = {arXiv},
	author = {Allen, Michael P. and Schmid, Friederike},
	month = jun,
	year = {2006},
	note = {arXiv:cond-mat/0606511}
}

@article{halonen_further_2023,
	title = {Further cautionary tales on thermostatting in molecular dynamics: {Energy} equipartitioning and non-equilibrium processes in gas-phase simulations},
	volume = {158},
	issn = {0021-9606, 1089-7690},
	shorttitle = {Further cautionary tales on thermostatting in molecular dynamics},
	url = {https://pubs.aip.org/jcp/article/158/19/194301/2890473/Further-cautionary-tales-on-thermostatting-in},
	doi = {10.1063/5.0148013},
	language = {en},
	number = {19},
	urldate = {2025-08-12},
	journal = {The Journal of Chemical Physics},
	author = {Halonen, Roope and Neefjes, Ivo and Reischl, Bernhard},
	month = may,
	year = {2023},
	pages = {194301}
}

@misc{irmer_granular_2024,
	title = {Granular temperature controls local rheology of vibrated granular flows},
	url = {http://arxiv.org/abs/2405.13236},
	doi = {10.48550/arXiv.2405.13236},
	language = {en},
	urldate = {2025-08-12},
	publisher = {arXiv},
	author = {Irmer, Mitchell G. and Brodsky, Emily E. and Clark, Abram H.},
	month = may,
	year = {2024},
	note = {arXiv:2405.13236 [cond-mat]}
}

@article{gdr_midi_dense_2004,
	title = {On dense granular flows},
	volume = {14},
	copyright = {http://www.springer.com/tdm},
	issn = {1292-8941, 1292-895X},
	url = {http://link.springer.com/10.1140/epje/i2003-10153-0},
	doi = {10.1140/epje/i2003-10153-0},
	language = {en},
	number = {4},
	urldate = {2025-08-12},
	journal = {The European Physical Journal E},
	author = {{GDR MiDi}},
	month = aug,
	year = {2004},
	pages = {341--365}
}

@article{pastorino2007comparison,
  title={Comparison of dissipative particle dynamics and Langevin thermostats for out-of-equilibrium simulations of polymeric systems},
  author={Pastorino, C and Kreer, T and M{\"u}ller, M and Binder, K},
  journal={Physical Review E—Statistical, Nonlinear, and Soft Matter Physics},
  volume={76},
  number={2},
  pages={026706},
  year={2007},
  publisher={APS}
}

@article{helfand1960transport,
  title={Transport coefficients from dissipation in a canonical ensemble},
  author={Helfand, Eugene},
  journal={Physical Review},
  volume={119},
  number={1},
  pages={1},
  year={1960},
  publisher={APS}
}

@article{degiuli2017friction,
  title={Friction law and hysteresis in granular materials},
  author={DeGiuli, E and Wyart, M},
  journal={Proceedings of the National Academy of Sciences},
  volume={114},
  number={35},
  pages={9284--9289},
  year={2017},
  publisher={National Academy of Sciences}
}

@article{tuckerman1999classical,
  title={On the classical statistical mechanics of non-Hamiltonian systems},
  author={Tuckerman, ME and Mundy, CJ and Martyna, GJ},
  journal={EPL (Europhysics Letters)},
  volume={45},
  number={2},
  pages={149--155},
  year={1999}
}

@article{garzo1999dense,
  title={Dense fluid transport for inelastic hard spheres},
  author={Garz{\'o}, V and Dufty, JW},
  journal={Physical Review E},
  volume={59},
  number={5},
  pages={5895},
  year={1999},
  publisher={APS}
}

@article{garzo2002transport,
  title={Transport coefficients of a heated granular gas},
  author={Garz{\'o}, Vicente and Montanero, Jos{\'e} Mar{\i}a},
  journal={Physica A: Statistical Mechanics and its Applications},
  volume={313},
  number={3-4},
  pages={336--356},
  year={2002},
  publisher={Elsevier}
}

@book{j2007statistical,
  title={Statistical mechanics of nonequilbrium liquids},
  author={J Evans, Denis and P Morriss, Gary},
  year={2007},
  publisher={ANU Press}
}

@article{campbell2002granular,
  title={Granular shear flows at the elastic limit},
  author={Campbell, Charles S},
  journal={Journal of fluid mechanics},
  volume={465},
  pages={261--291},
  year={2002},
  publisher={Cambridge University Press}
}

@article{ogarko2012equation,
  title={Equation of state and jamming density for equivalent bi-and polydisperse, smooth, hard sphere systems},
  author={Ogarko, V and Luding, Stefan},
  journal={The Journal of chemical physics},
  volume={136},
  number={12},
  year={2012},
  publisher={AIP Publishing}
}

@article{hoover2004time,
  title={Time-reversible deterministic thermostats},
  author={Hoover, Wm G and Aoki, Kenichiro and Hoover, Carol G and De Groot, Stephanie V},
  journal={Physica D: Nonlinear Phenomena},
  volume={187},
  number={1-4},
  pages={253--267},
  year={2004},
  publisher={Elsevier}
}

@article{visco2007power,
  title={Power injected in a granular gas},
  author={Visco, Paolo and Puglisi, Andrea and Barrat, Alain and Trizac, Emmanuel and van Wijland, Fr{\'e}d{\'e}ric},
  journal={Comptes Rendus Physique},
  volume={8},
  number={5-6},
  pages={641--649},
  year={2007},
  publisher={Elsevier}
}

@article{montanero2000computer,
  title={Computer simulation of uniformly heated granular fluids},
  author={Montanero, Jos{\'e} Maria and Santos, Andr{\'e}s},
  journal={Granular Matter},
  volume={2},
  number={2},
  pages={53--64},
  year={2000},
  publisher={Springer}
}

@incollection{brilliantov2001granularb,
  title={Granular gases with impact-velocity-dependent restitution coefficient},
  author={Brilliantov, Nikolai V and P{\"o}schel, Thorsten},
  booktitle={Granular Gases},
  pages={100--124},
  year={2001},
  publisher={Springer}
}

@inproceedings{brilliantov2001granular,
  title={Granular gases—The early stage},
  author={Brilliantov, Nikolai V and P{\"o}schel, Thorsten},
  booktitle={Coherent Structures in Complex Systems: Selected Papers of the XVII Sitges Conference on Statistical Mechanics Held a Sitges, Barcelona, Spain, 5--9 June 2000},
  pages={408--419},
  year={2001},
  organization={Springer}
}

@article{garzo2004diffusion,
  title={Diffusion of impurities in a granular gas},
  author={Garz{\'o}, Vicente and Montanero, Jos{\'e} Mar{\'\i}a},
  journal={Physical Review E},
  volume={69},
  number={2},
  pages={021301},
  year={2004},
  publisher={APS}
}

@article{gomart2004granular,
  title={Granular discorectangle in a thermalized bath of hard disks},
  author={Gomart, H and Talbot, J and Viot, P},
  journal={arXiv preprint cond-mat/0408205},
  year={2004}
}

@article{fullmer2022divergence,
  title={The divergence of nearby trajectories in soft-sphere DEM},
  author={Fullmer, William D and Porcu, Roberto and Musser, Jordan and Almgren, Ann S and Srivastava, Ishan},
  journal={Particuology},
  volume={63},
  pages={1--8},
  year={2022},
  publisher={Elsevier}
}

@article{kang2010granular,
  title={Granular gases under extreme driving},
  author={Kang, W and Machta, Jonathan and Ben-Naim, E},
  journal={EPL (Europhysics Letters)},
  volume={91},
  number={3},
  pages={34002},
  year={2010}
}

@article{shah2025molecular,
  title={Molecular dynamics study of diffusion coefficient in Uniformly Heated Hard Sphere Granular Gas in three dimensions},
  author={Shah, Rameez Farooq and Ahmad, Syed Rashid},
  journal={Physica A: Statistical Mechanics and its Applications},
  volume={657},
  pages={130221},
  year={2025},
  publisher={Elsevier}
}

@article{berzi2024granular,
  title={On granular flows: From kinetic theory to inertial rheology and nonlocal constitutive models},
  author={Berzi, Diego},
  journal={Physical Review Fluids},
  volume={9},
  number={3},
  pages={034304},
  year={2024},
  publisher={APS}
}

@article{viscardy2007transport,
  title={Transport and Helfand moments in the Lennard-Jones fluid. I. Shear viscosity},
  author={Viscardy, S{\'e}bastien and Servantie, James and Gaspard, Pierre},
  journal={The Journal of chemical physics},
  volume={126},
  number={18},
  year={2007},
  publisher={AIP Publishing}
}

@book{poschel2005computational,
  title={Computational granular dynamics: models and algorithms},
  author={P{\"o}schel, Thorsten and Schwager, Thomas},
  year={2005},
  publisher={Springer}
}

@article{knight1995density,
  title={Density relaxation in a vibrated granular material},
  author={Knight, James B and Fandrich, Christopher G and Lau, Chun Ning and Jaeger, Heinrich M and Nagel, Sidney R},
  journal={Physical review E},
  volume={51},
  number={5},
  pages={3957},
  year={1995},
  publisher={APS}
}

@article{goldhirsch2000green,
  title={Green-Kubo relations for granular fluids},
  author={Goldhirsch, I and Van Noije, TPC},
  journal={Physical Review E},
  volume={61},
  number={3},
  pages={3241},
  year={2000},
  publisher={APS}
}

@article{mandal2021rheology,
  title={Rheology of cohesive granular media: shear banding, hysteresis, and nonlocal effects},
  author={Mandal, Sandip and Nicolas, Maxime and Pouliquen, Olivier},
  journal={Physical Review X},
  volume={11},
  number={2},
  pages={021017},
  year={2021},
  publisher={APS}
}

@article{bai2019crystallization,
  title={Crystallization via shaking in a granular gas with van der Waals interactions},
  author={Bai, Qiong and Mazza, Marco G},
  journal={Physical Review E},
  volume={100},
  number={4},
  pages={042910},
  year={2019},
  publisher={APS}
}

@article{carnahan1969equation,
  title={Equation of state for nonattracting rigid spheres},
  author={Carnahan, Norman F and Starling, Kenneth E},
  journal={Journal of chemical physics},
  volume={51},
  number={2},
  pages={635--636},
  year={1969}
}

@article{vescovi2016merging,
  title={Merging fluid and solid granular behavior},
  author={Vescovi, Dalila and Luding, Stefan},
  journal={Soft matter},
  volume={12},
  number={41},
  pages={8616--8628},
  year={2016},
  publisher={Royal Society of Chemistry}
}

@article{stoyanov_molecular_2005,
	title = {From molecular dynamics to hydrodynamics: {A} novel {Galilean} invariant thermostat},
	volume = {122},
	issn = {0021-9606, 1089-7690},
	shorttitle = {From molecular dynamics to hydrodynamics},
	url = {https://pubs.aip.org/jcp/article/122/11/114112/929978/From-molecular-dynamics-to-hydrodynamics-A-novel},
	doi = {10.1063/1.1870892},
	language = {en},
	number = {11},
	urldate = {2025-08-12},
	journal = {The Journal of Chemical Physics},
	author = {Stoyanov, Simeon D. and Groot, Robert D.},
	month = mar,
	year = {2005},
	pages = {114112}
}

@article{komatsu2015roles,
  title={Roles of energy dissipation in a liquid-solid transition of out-of-equilibrium systems},
  author={Komatsu, Yuta and Tanaka, Hajime},
  journal={Physical Review X},
  volume={5},
  number={3},
  pages={031025},
  year={2015},
  publisher={APS}
}

@article{bussi_canonical_2007,
	title = {Canonical sampling through velocity-rescaling},
	volume = {126},
	issn = {0021-9606, 1089-7690},
	url = {http://arxiv.org/abs/0803.4060},
	doi = {10.1063/1.2408420},
	language = {en},
	number = {1},
	urldate = {2025-08-12},
	journal = {The Journal of Chemical Physics},
	author = {Bussi, Giovanni and Donadio, Davide and Parrinello, Michele},
	month = jan,
	year = {2007},
	note = {arXiv:0803.4060 [cond-mat]},
	pages = {014101}
}

@article{verbeek_advantages_2022,
	title = {Advantages of the {Rayleigh}–{Lowe}–{Andersen} thermostat in soft sphere molecular dynamics simulations},
	volume = {45},
	issn = {1292-8941, 1292-895X},
	url = {https://link.springer.com/10.1140/epje/s10189-022-00173-7},
	doi = {10.1140/epje/s10189-022-00173-7},
	language = {en},
	number = {3},
	urldate = {2025-08-12},
	journal = {The European Physical Journal E},
	author = {Verbeek, Martijn G. and Smid, Dietha and Valentijn, Jasper and Valentijn, Joop},
	month = mar,
	year = {2022},
	pages = {27}
}

@article{hoogerbrugge1992simulating,
  title={Simulating microscopic hydrodynamic phenomena with dissipative particle dynamics},
  author={Hoogerbrugge, PJ and Koelman, Johannes MVA},
  journal={Europhysics letters},
  volume={19},
  number={3},
  pages={155},
  year={1992},
  publisher={IOP Publishing}
}

@article{nose1984unified,
  title={A unified formulation of the constant temperature molecular dynamics methods.},
  author={Nosé, Shuichi},
  journal={The Journal of Chemical Physics},
  volume={81},
  pages={511},
  year={1984}
}

@article{langevin1908theory,
  title={On the theory of brownian motion.},
  author={Langevin, Paul},
  journal={CR Acad Sci (Paris)},
  volume={146},
  pages={530},
  year={1908}
}

@article{ruiz2018effect,
  title={On the effect of the thermostat in non-equilibrium molecular dynamics simulations},
  author={Ruiz-Franco, Jos{\'e} and Rovigatti, Lorenzo and Zaccarelli, Emanuela},
  journal={The European Physical Journal E},
  volume={41},
  number={7},
  pages={80},
  year={2018},
  publisher={Springer}
}

@article{jop2006constitutive,
  title={A constitutive law for dense granular flows},
  author={Jop, Pierre and Forterre, Yo{\"e}l and Pouliquen, Olivier},
  journal={Nature},
  volume={441},
  number={7094},
  pages={727--730},
  year={2006},
  publisher={Nature Publishing Group UK London}
}

@article{plati2021getting,
  title={Getting hotter by heating less: How driven granular materials dissipate energy in excess},
  author={Plati, A and De Arcangelis, L and Gnoli, A and Lippiello, E and Puglisi, A and Sarracino, A},
  journal={Physical Review Research},
  volume={3},
  number={1},
  pages={013011},
  year={2021},
  publisher={APS}
}

@article{dumont2023microscopic,
  title={Microscopic foundation of the $\mu$ (I) rheology for dense granular flows on inclined planes},
  author={Dumont, Denis and Bonneau, Haggai and Salez, Thomas and Raphael, Elie and Damman, Pascal},
  journal={Physical Review Research},
  volume={5},
  number={1},
  pages={013089},
  year={2023},
  publisher={APS}
}

@article{wang2025basal,
  title={Basal layer of granular flow down smooth and rough inclines: kinematics, slip laws and rheology},
  author={Wang, Teng and Jing, Lu and Kwok, CY and Sobral, Yuri D and Weinhart, Thomas and Thornton, Anthony R},
  journal={Journal of Fluid Mechanics},
  volume={1025},
  pages={A27},
  year={2025},
  publisher={Cambridge University Press}
}

@article{windows2013boltzmann,
  title={Boltzmann statistics in a three-dimensional vibrofluidized granular bed: Idealizing the experimental system},
  author={Windows-Yule, CRK and Parker, DJ},
  journal={Physical Review E—Statistical, Nonlinear, and Soft Matter Physics},
  volume={87},
  number={2},
  pages={022211},
  year={2013},
  publisher={APS}
}

@article{berzi2014extended,
  title={Extended kinetic theory applied to dense, granular, simple shear flows},
  author={Berzi, Diego},
  journal={Acta Mechanica},
  volume={225},
  number={8},
  pages={2191--2198},
  year={2014},
  publisher={Springer}
}

\end{document}